\documentclass[onefignum,leqno]{siamltex}   

\usepackage{epsfig} 
\usepackage{graphicx}
\usepackage{dcolumn}
\usepackage{bm}
\usepackage[usenames]{color}
\usepackage{amsmath}
\usepackage{amssymb}

\title{Escape Rates in a Stochastic Environment with Multiple Scales}

\author{Eric Forgoston\thanks{Nonlinear Dynamical Systems Section, Plasma
    Physics Division, Code 6792, U.S. Naval Research Laboratory, Washington,
    DC 20375, USA ({\tt eric.forgoston.ctr@nrl.navy.mil}).} \and Ira
  B. Schwartz\thanks{Nonlinear Dynamical Systems Section, Plasma
    Physics Division, Code 6792, U.S. Naval Research Laboratory, Washington,
    DC 20375, USA ({\tt ira.schwartz@nrl.navy.mil}).}}

\begin{document}
\maketitle

\begin{abstract}
We consider a stochastic
environment with two  time scales and outline a general theory
that compares two methods to reduce the dimension of the original system.  The
first method involves the computation of the underlying deterministic center
manifold followed by a ``na{\"i}ve'' replacement of the stochastic term.  The second
method allows one to more accurately describe the stochastic effects and
involves the derivation of a normal form coordinate transform that is used to
find the stochastic center manifold.  The results of both methods
are used along with the path integral formalism of large fluctuation
theory to predict the escape
rate from one basin of attraction to another.  The general theory is applied to
the example of a surface flow described
by a generic, singularly perturbed, damped, nonlinear oscillator with additive,
Gaussian noise.  We show how both nonlinear reduction methods compare in
escape rate scaling.  Additionally, the center manifolds are shown to predict
high pre-history probability regions of escape.  The theoretical results are
confirmed using numerical computation of the mean escape time and escape
prehistory, and we briefly discuss the extension of the theory to stochastic control.
\end{abstract}



\pagestyle{myheadings}
\thispagestyle{plain}
\markboth{E. FORGOSTON AND IRA B. SCHWARTZ}{ESCAPE RATES IN A STOCHASTIC ENVIRONMENT}

\section{Introduction}

It has long been known that noise can have a significant effect on deterministic dynamical
systems.  For example, given an initial state starting in some basin of
attraction (defined as the set of initial conditions 
from which
the system approaches a corresponding locally stable attractor as time evolves
to infinity), noise can cause the initial state to cross the basin boundary
and move into
another, distinct basin of attraction~\cite{dyk90,dmsss92,mil96,lumcdy98,llbs03}.  

There are several points of view one might consider when investigating the
effect of noise on a dynamical system,
including stochastic resonance~\cite{ghjm98} and finite noise
effects \cite{bbs02}.  In this article, we consider yet another point of view,
namely the effect of arbitrarily small noise on the escape of a particle from
a potential well.  In this case, one can apply
large fluctuation theory~\cite{feyhib65,dyk90,dmsss92,lumcdy98}.  

Many of the
underlying deterministic systems found in~\cite{dyk90,dmsss92,mil96,lumcdy98,llbs03} have parameter
regimes in which multiple attractors give rise to
noise-induced escape from one attractor to another. Such systems
may be analyzed globally by considering the Hamiltonian theory
of large fluctuations or by considering escape from attracting
potential wells along most probable exit paths~\cite{FW84,GT84,MS93,HTG94}.

Through the use of a path
integral coupled with variational methods, it is possible to compute the probability
densities of the trajectories of the system.  In particular, for sufficiently small noise, one can find the trajectories which escape from
a basin of attraction due to stochastic effects.  The most probable escape trajectory is
the optimal escape path of a state residing in a basin of attraction.

Many researchers have
investigated how noise affects physical and biological phenomena, including
lasers~\cite{cbst00,hwgali00,hzrd00},
epidemics and control~\cite{BillingsBS02,dyscsh05,kammee08,dyscla08}, and neurons~\cite{hjbs06}.
Yet another important application in many fields is that of sensing
in stochastic environments.  Improved environmental sensing and prediction can
be achieved 
through the incorporation of continuous
monitoring of the region of interest.  For example, one could monitor the
stochastic ocean using autonomous underwater
gliders~\cite{webbsijo01,shermandov01,eriksenolwlsbc01}.  However, to do this,
one must understand both the dynamics and control of the gliders.

Extending the lifetime (energy optimization problem) of sensing devices
(e.g. gliders) in stochastic environments such as the ocean requires
an understanding of the effect of the environmental forces on both the devices
and the
region being monitored.  The ocean dynamics are high-dimensional and stochastic.  Therefore, as a first step towards using the underlying ocean structure to optimize a
sensor's energy usage, 
we will outline a general theory that provides two
methods to obtain a reduction in the dimension of the stochastic system.  The manifold
equations that are found using these methods can then be used
to determine the optimal escape path and escape rate. 

Our formulation uses large fluctuation theory~\cite{feyhib65} to determine the first passage times in a multi-scale
environment.  For a
vector field that has relaxation times on the same scale, it is clear how
to use the theory to generate an optimal path of escape, and this theory has
been applied to a variety of Hamiltonian and Lagrangian variational problems~\cite{wentzell76,Hu1987,Dykman1994d}. 

However, 
technical issues may arise when one wishes to determine the projection of
noise that is needed to
perturb the dynamics restricted to the lower-dimensional manifold. To address
this, several approaches have been developed to understand dimension reduction in
systems that have well separated time scales.  For a system with certain spectral
requirements, the existence of a stochastic center
manifold was proven in~\cite{box89}.  Non-rigorous stochastic normal form analysis (which leads to the
stochastic center manifold) was performed
in~\cite{knowie83,coelti85,nam90,namlin91}.  Rigorous theoretical analysis of
normal form coordinate transformations for stochastic center manifold reduction
was developed in~\cite{arnimk98,arn98}.  Later, an alternative method of stochastic
normal form reduction was developed~\cite{rob08}, in which any
anticipatory convolutions (integrals into the future of the noise processes) that appeared in the slow modes were
removed. Since this latter analysis makes the construction of the stochastic normal
form coordinate transform more
transparent, we use this method to derive the reduced stochastic center
manifold equation. 

The layout of the paper is as follows. The general theory of deterministic and
stochastic center manifold reduction is described in Sec.~\ref{sec:gen_theory}.  The first method used to reduce
the dimension of the system
involves the derivation of the center manifold equation~\cite{car81} of
the associated deterministic system followed by the ``na{\"i}ve'' replacement
of the stochastic term.  The second method, which allows one to more accurately describe
the effect of the noise, involves the derivation of a normal form
coordinate transform~\cite{rob08} that is used to find the stochastic center
manifold
equation.  Section~\ref{sec:gen_theory} also describes how the two center manifolds resulting
from the two methods can be used
along with the
theory of large fluctuations to analytically find the optimal escape path
of the particle along with its escape rate.  The general theory of Sec.~\ref{sec:gen_theory} is applied to a specific
example given by a singularly perturbed, damped, Duffing oscillator with
additive, Gaussian noise in Sec.~\ref{sec:example}.  This section contains analytical
results and their comparison with numerical computation.  The conclusions are
contained in Sec.~\ref{sec:conc}.

\section{General Theory} \label{sec:gen_theory}

We consider the following general $(m+n)$-dimensional system of stochastic differential equations with two
well-separated time scales:\clearpage
\begin{subequations}
\begin{equation}
\label{e:Gen_sys_x}
\dot{\bf{x}}={\bf A}{\bf x} + {\bf F}\left ({\bf x},{\bf y},{\bm \Phi}\right ),
\end{equation}  
\begin{equation}
\label{e:Gen_sys_y}
\epsilon\,\dot{\bf{y}}={\bf B}{\bf y} + {\bf G}\left ({\bf x},{\bf y},{\bm
    \Psi}\right ),
\end{equation}  
\end{subequations}
where $\epsilon$ is a small parameter, ${\bf x}(t)\in\mathbb{R}^m$, ${\bf
  y}(t)\in\mathbb{R}^n$, ${\bm \Phi}(t)$ and ${\bm \Psi}(t)$ describe stochastic forces with
adjustable intensity, ${\bf A}$ and ${\bf B}$ are constant matrices, and
${\bf F}$ and ${\bf G}$ are stochastic, nonlinear functions.

\subsection{Deterministic Center Manifold}\label{sec:gen_theory_DCM}

To begin, we remove the stochastic terms from Eqs.~(\ref{e:Gen_sys_x})
and~(\ref{e:Gen_sys_y}) so that ${\bf F}={\bf F}({\bf x},{\bf y})$ and ${\bf
  G}={\bf G}({\bf x},{\bf y})$.  Let $t=\epsilon\tau$.  Denoting $\dot{}$ as
$d/dt$ and $^{\prime}$ as $d/d\tau$, then the deterministic form of
Eqs.~(\ref{e:Gen_sys_x}) and~(\ref{e:Gen_sys_y}) is transformed to the
following system of equations: 
\begin{subequations}
\begin{equation}
\label{e:trans_sys_x}
{\bf{x}}^{\prime}=\epsilon\left [{\bf A}{\bf x} + {\bf F}({\bf x},{\bf
    y})\right ],
\end{equation}  
\begin{equation}
\label{e:trans_sys_y}
{\bf{y}}^{\prime}={\bf B}{\bf y} + {\bf G}({\bf x},{\bf y}),
\end{equation}
\begin{equation}
\label{e:trans_sys_eps}
\epsilon^{\prime}=0.
\end{equation}  
\end{subequations}

To recast the problem in a more general framework, we treat $\epsilon$ as
  a state variable, let ${\bf \bar{A}}=\epsilon {\bf A}$ and ${\bf
    \bar{F}}=\epsilon {\bf F}$, and write Eqs.~(\ref{e:trans_sys_x})-(\ref{e:trans_sys_eps}) as
\begin{subequations}
\begin{equation}
\label{e:rewrit_x}
{\bf{x}}^{\prime}={\bf \bar{A}}{\bf x} + {\bf \bar{F}}({\bf x},{\bf y},\epsilon),
\end{equation}
\begin{equation}
\label{e:rewrit_y}
{\bf{y}}^{\prime}={\bf B}{\bf y} + {\bf G}({\bf x},{\bf y}),
\end{equation}
\begin{equation}
\label{e:rewrit_eps}
\epsilon^{\prime}=0.
\end{equation}  
\end{subequations}
If ${\bf \bar{A}}$ and ${\bf B}$ are constant matrices such that all
of the eigenvalues of ${\bf \bar{A}}$ have zero real parts, while all of the
eigenvalues of ${\bf B}$ have negative real parts, then the system will
rapidly collapse onto a lower-dimensional manifold given by center manifold
theory~\cite{car81}.  Furthermore, we will consider examples where the
  solution decays throughout the transient and then stays close to the
  lower-dimensional manifold.

If the center manifold is given by 
\begin{equation}
\label{e:cent_man}
{\bf y}={\bf h}({\bf x},\epsilon),
\end{equation}
then substitution of Eq.~(\ref{e:cent_man}) into Eq.~(\ref{e:rewrit_y}) leads
to the following center manifold condition:
\begin{equation}
\label{e:cent_man_cond}
{\bf h}_{{\bf x}}\left [{\bf \bar{A}}{\bf x} + {\bf \bar{F}}({\bf
  x},{\bf h}({\bf x},\epsilon),\epsilon)\right ] = {\bf B}{\bf h}({\bf
  x},\epsilon) + {\bf G}({\bf
  x},{\bf h}({\bf x},\epsilon)),
\end{equation}
where ${\bf h}_{{\bf x}}$ denotes the partial derivative  of ${\bf h}$ with
respect to ${\bf x}$.  Although it is
generally not possible to solve Eq.~(\ref{e:cent_man_cond}) for ${\bf h}$, one can
approximate the center manifold by expanding ${\bf h}$ in the following way:
\begin{equation}
\label{e:expand}
{\bf h}({\bf x},\epsilon)={\bf h_0}({\bf x})+\epsilon {\bf
  h_1}({\bf x})+\epsilon^2 {\bf h_2}({\bf x})+\mathcal{O}(\epsilon^3).
\end{equation}
Typically, this approximation of ${\bf h}({\bf x},\epsilon)$ is found by
substituting Eq.~(\ref{e:expand}) into the center manifold condition
[Eq.~(\ref{e:cent_man_cond})] and matching coefficients.
\subsection{Optimal Escape Path and Escape Rate}\label{sec:gen_theory_OEP}
Starting with the center manifold equation given by Eq.~(\ref{e:expand}) to a
particular order, the dynamics on the center manifold can be found by
substitution of Eq.~(\ref{e:expand}) into Eq.~(\ref{e:trans_sys_x}) [using the
relation given by Eq.~(\ref{e:cent_man})].  Therefore, the dynamics are
determined by
\begin{equation}
\label{e:dyn}
{\bf{x}}^{\prime}=\epsilon \left [{\bf A}{\bf x} + {\bf F}({\bf
  x},{\bf h}({\bf x},\epsilon))\right ]=\epsilon {\bf H}({\bf x},\epsilon),
\end{equation}
and use of the relation between $t$ and $\tau$ leads to the following:
\begin{equation}
\label{e:dyn_2}
\dot{{\bf{x}}}={\bf A}{\bf x} + {\bf F}({\bf
  x},{\bf h}({\bf x},\epsilon))={\bf H}({\bf x},\epsilon).
\end{equation}

Equation~(\ref{e:dyn_2}) is a deterministic equation.  However, now that we have
reduced the dimension of the problem, we return to considering a stochastic
problem by ``na{\"i}vely'' adding a noise vector to the right-hand side of
Eq.~(\ref{e:dyn_2}) so that one has
\begin{equation}
\label{e:dyn_stoch}
\dot{{\bf{x}}}={\bf H}({\bf x},\epsilon)+\sqrt{2D}{\bm \Phi}(t),
\end{equation}
where $D$ is the noise intensity.  Each of the noise components, $\phi_i$, of ${\bm \Phi}$ in Eq.~(\ref{e:dyn_stoch})
describes a stochastic white force that is
characterized by the following correlation functions:
\begin{subequations}
\begin{equation}
\label{e:mean}
\langle\phi_i(t)\rangle=0,
\end{equation}
\begin{equation}
\label{e:corr}
\langle\phi_i(t)\phi_j(t^{\prime})\rangle=\delta(t-t^{\prime})\delta_{ij}.
\end{equation}
\end{subequations}

The following analysis to determine the optimal escape path may be
performed using Eqs.~(\ref{e:dyn_stoch})-(\ref{e:corr}), a system with an
arbitrary number of degrees of freedom~\cite{einmck96}.  However, since we
ultimately are interested in applying this general theory to a two-dimensional (2D)
surface flow that reduces via the center manifold to a one-dimensional (1D)
equation, for simplicity we consider the 1D version of Eq.~(\ref{e:dyn_stoch})
given as  
\begin{equation}
\label{e:dyn_stoch_1D}
\dot{x}=H(x,\epsilon)+\sqrt{2D}\phi (t),
\end{equation}
where $\phi (t)$ is characterized by the correlation functions given by
Eqs.~(\ref{e:mean}) and (\ref{e:corr}).

We assume that $H(x,\epsilon)$ is associated with a potential function
$U(x,\epsilon)$, \linebreak $[H(x,\epsilon)=-dU(x,\epsilon)/dx]$ with stable states (attractors) located at $x=x_a$
and an unstable state (saddle) located at $x=x_s$.  Then
Eq.~(\ref{e:dyn_stoch_1D}) corresponds to a Langevin equation of a particle in
an over-damped potential well.

In the absence of noise, a particle located in the potential well will
approach the stable, attracting state.  However, the noise may organize as an
effective force which acts on the particle and ``pushes'' the particle from
the attractor to the saddle located at the top of the potential well barrier.
The path along which the particle leaves the basin of attraction due to such
an effective noise force is an escape path.

Given that $\phi(t)$ is uncorrelated Gaussian noise,
the probability of optimal escape is
\begin{equation}
\label{e:opt_esc}
P[x_{\rm esc}]=C\exp{\left (-\frac{1}{2D}\int\limits_{-\infty}^{\infty} \phi_{\rm
  opt}^2 \,\,dt\right )}=C\exp{(-R/D)},
\end{equation}
where
\begin{equation}
\label{e:R}
R=\frac{1}{2}\int\limits_{-\infty}^{\infty} \phi_{\rm opt}^2\,\,dt=\frac{1}{2}\int\limits_{-\infty}^{\infty}L(x,\dot{x};t)\,\,dt,
\end{equation}
and $\phi_{\rm opt}$ denotes the stochastic fluctuations corresponding to the
trajectory that moves along the optimal escape path.  In
Eq.~(\ref{e:opt_esc}), $C$ is a pre-factor that depends on the noise
intensity.  In order to maximize the
probability of escape, one must minimize the exponent given
by Eq.~(\ref{e:R}).  Using the Euler-Lagrange
equation of motion, it is now possible to solve for the optimal escape
path.  

It also is possible to derive an expression for the escape rate from an attractor to the
saddle, which is located at the top of the potential well barrier.  In general, if the stochastic trajectory of a particle is given by
\begin{equation}
\label{e:U_eq}
\dot{x}=-\frac{dU(x,\epsilon)}{dx}+\sqrt{2D}\phi(t),
\end{equation}
with $U(x,\epsilon)$ some potential (as we have assumed), then for Gaussian noise, the escape rate from the
attractor (located at $x=x_a$) to the unstable saddle (located at $x=x_s$ at
the top of the barrier) is
given by
\begin{equation}
\label{e:W}
W(D)=\frac{\sqrt{|Q(x_s,\epsilon)|Q(x_a,\epsilon)}}{2\pi}\exp{(-\Delta U/D)},
\end{equation}
where $Q(x,\epsilon)=d^2 U(x,\epsilon)/dx^2$ and $\Delta U$ is the activation energy of escape (depth of the potential
well)~\cite{gar03}.

If $W(D)$ is the escape rate, then $1/W(D)$ gives the mean escape time, and
the natural log of the mean escape time is therefore
\begin{equation}
\label{e:log-MET}
\log{\left (\frac{1}{W(D)}\right )}=\log{\left
    (\frac{2\pi}{\sqrt{|Q(x_s,\epsilon)|Q(x_a,\epsilon)}}\right )}
+ \frac{\Delta U}{D}.
\end{equation}

It should be noted that to ensure the particle escapes from the
potential well, one could compute the escape rate (and the mean escape
time) from the attractor to a point located somewhere in the second
potential well (i.e. the particle escapes from the first basin of attraction if it climbs
out of the potential well to the saddle located at the top of the potential well barrier
and then continues past the saddle to some point located in the second basin of attraction).
However, this movement of the point at which one claims the particle has escaped from the
potential well will create a change in the pre-factor of Eq.~(\ref{e:W}) along
with a corresponding change in the first term on the right-hand side of
Eq.~(\ref{e:log-MET})~\cite{gar03}. 

As we have previously noted, we consider systems whose solution decays
exponentially throughout the transient and then stays close to the lower-dimensional
center manifold.  There are no secular terms in the asymptotic expansion since we are not looking at periodic
orbits, and the result is valid for all time.  Moreover, any noise drift on
the center manifold will result in bounded solutions due to sufficient
dissipation transverse to the manifold.  This behavior is in direct contrast to the finite-time solutions
on manifolds of relaxation-type oscillators whereby the time scale of escape
must be shorter than the lifetime of the trajectories on the manifold~\cite{dv-em05,fre01}.  This
lifetime issue does not pertain to the systems we consider since there will be
no oscillations.

\subsection{Stochastic Center Manifold and the Normal Form Coordinate Transform}\label{sec:gen_theory_SCM}

In a manner similar to that shown in Sec.~\ref{sec:gen_theory_DCM}, the stochastic system given by Eqs.~(\ref{e:Gen_sys_x})
and~(\ref{e:Gen_sys_y}) can be transformed~\cite{bergen03} to the form
given by Eqs.~(\ref{e:rewrit_x})-(\ref{e:rewrit_eps}), where now ${\bf
  \bar{F}}={\bf \bar{F}}({\bf x},{\bf y},\epsilon,{\bm \Phi})$ and ${\bf
  G}={\bf G}({\bf x},{\bf y},{\bm \Psi})$.  If ${\bf
  \bar{A}}$ and ${\bf B}$ satisfy the same spectral conditions as for
the deterministic system, and if the stochastic time dependence found in ${\bf
  \bar{F}}$ and ${\bf
  G}$ is due to independent white noise processes, then there exists a
stochastic center manifold for the original stochastic system~\cite{box89}. 

One method for computing the stochastic center manifold for systems with
both fast and slow dynamics uses the construction of a normal form coordinate
transform that not only reduces the dimension of the dynamics, but also
separates all of the fast processes from all of the slow
processes~\cite{rob08}.  While this type of normal form coordinate transform
may be used to find deterministic center manifolds, the application of this
transform to stochastic systems is particularly interesting since white noise
has fluctuations on all scales.

There are many publications, such as~\cite{knowie83,coelti85,nam90,namlin91} which deal with the simplification of a stochastic
dynamical system using a stochastic normal form transformation.  In these
articles, the noise term is multiplied by a small parameter, and therefore,
the resulting stochastic normal form is a perturbation of the deterministic
normal form.  Furthermore, one can find in~\cite{coelti85,namlin91} normal
form transformations that involve anticipative noise processes.  However,
these integrals of the noise process into the future were not dealt with
rigorously in~\cite{coelti85,namlin91}. 

Rigorous, theoretical analysis to support normal form coordinate transforms
(and center manifold reduction) was developed in~\cite{arnimk98,arn98}.  In
this work, the technical problem of the anticipative noise integrals also was
dealt with rigorously.  Later, another stochastic normal form transformation
was developed~\cite{rob08}.  This new method is such that ``anticipation can ... always [be] removed from
the slow modes with the result that no anticipation is required after the fast
transients decay''(Ref.~\cite{rob08}, pp. 13).  An advantage of removing
anticipation is the simplification of the normal form.  Nonetheless, this
simpler normal form retains its accuracy with the original stochastic system.
Furthermore, when modeling the macroscopic behavior of microscopic,
stochastic systems, it is desirable to avoid anticipation in the normal form~\cite{rob08}. 

It is important to note that the normal form is valid for all time since it is
just a coordinate transform.  Furthermore, the dynamics also are valid for all
time as long as the truncation error is small enough for the problem of interest.

In the example of Sec.~\ref{sec:example}, we shall use the method of~\cite{rob08} to
simplify our stochastic dynamical system to one that emulates the long-term dynamics of
the original, multiple-time-scale system.  The method involves five
principles, which we recapitulate here for the purpose of clarity.  The
principles are as follows:
\begin{remunerate}
\item Avoid unbounded, secular terms in both the transformation and the
  evolution equations to ensure a uniform asymptotic approximation.\label{one}
\item Decouple all of the slow processes from the fast processes to ensure a
  valid long-term model.\label{two}
\item Insist that the stochastic slow manifold is precisely the transformed
  fast processes coordinate being equal to zero.\label{three}
\item To simplify matters, eliminate as many as possible of the terms in the evolution equations.\label{four}
\item Try to remove all fast processes from the slow processes by avoiding as
  much as possible the fast time memory integrals in the evolution equations.\label{five}
\end{remunerate}

In practice, the original stochastic system of equations (which satisfy the
necessary spectral requirements) in $({\bf x},{\bf y})$ coordinates is
transformed to a new $({\bf X},{\bf Y})$ coordinate system using a stochastic
coordinate transform as follows:
\begin{subequations}
\begin{equation}
\label{e:stoch_coord_trans_x}
{\bf x}={\bf X}+{\bm \xi}({\bf X},{\bf Y},t),
\end{equation}
\begin{equation}
\label{e:stoch_coord_trans_y}
{\bf y}={\bf Y}+{\bm \eta}({\bf X},{\bf Y},t),
\end{equation}
\end{subequations}
where the specific form of Eqs.~(\ref{e:stoch_coord_trans_x}) and~(\ref{e:stoch_coord_trans_y}) is chosen to simplify the original
system according to the five principles listed previously.  The terms ${\bm
  \xi}({\bf X},{\bf Y},t)$ and ${\bm \eta}({\bf X},{\bf Y},t)$ are found using
an iterative procedure that will be demonstrated using the singularly
perturbed, damped, stochastic Duffing oscillator model in Sec.~\ref{sec:example}.  Theoretical
details can be found in~\cite{rob08}.

\section{Example - Singularly Perturbed Stochastic Duffing Oscillator}\label{sec:example}

We consider the following singularly perturbed, damped, Duffing oscillator
system with additive noise:
\begin{subequations}
\begin{equation}
\label{e:Duff_sys_x}
\dot{x}=y + \sqrt{2D}\phi(t),
\end{equation}
\begin{equation}
\label{e:Duff_sys_y}
\epsilon\dot{y}=(x-x^3-y),
\end{equation}
\end{subequations}
where $D$ is the noise intensity and $\phi(t)$ describes a stochastic white
force that is characterized by the correlation
functions given in Eqs.~(\ref{e:mean}) and~(\ref{e:corr}).  

In this example, the noise is added only to the $x$ equation.  Additionally,
one could consider two other scenarios.  In the first scenario, noise is added
to both the $x$ and $y$ equations, while in the second scenario, noise is
added only to the $y$ equation.  To implement the first scenario, one
adds $\sqrt{2D}\phi_1(t)$ to the right-hand side of the $x$ equation and 
$\sqrt{2D}\phi_2(t)$ to the right-hand side of the $y$ equation [where
$\phi_1(t)$ and $\phi_2(t)$ describe stochastic white
forces of intensity $D$ that are characterized by the correlation
functions given in Eqs.~(\ref{e:mean}) and~(\ref{e:corr})].  To implement the second scenario, noise is
added only to the $y$ equation.  However, unlike the example
[Eqs.~(\ref{e:Duff_sys_x}) and (\ref{e:Duff_sys_y})] and the first scenario, in
this case the noise term must be scaled by $\sqrt{\epsilon}$ and the potential
function must be scaled by $\epsilon$~\cite{schspi98}.
Although the following results pertain to Eqs.~(\ref{e:Duff_sys_x})
and~(\ref{e:Duff_sys_y}), we have checked that one obtains similar results for the other two scenarios.

The system given by Eqs.~(\ref{e:Duff_sys_x})
and~(\ref{e:Duff_sys_y}) is very strongly damped when $\epsilon
\ll 1$.  For the case of an under-damped system, one should consider the dynamics in the limit of
weak damping~\cite{dyscsh05}.

\subsection{Deterministic Center Manifold}\label{sec:example_DCM}

Following the general theory of Sec. \ref{sec:gen_theory_DCM}, we consider the deterministic form
of Eqs.~(\ref{e:Duff_sys_x}) and~(\ref{e:Duff_sys_y}) by setting $\phi(t)=0$.  The slow manifold is found by setting $\epsilon =0$ in Eq.~(\ref{e:Duff_sys_y}).
Solving for $y$ gives the equation of the slow manifold as $y=x-x^3$[which
corresponds to $h_0(x)$ in Eq.~(\ref{e:expand})].
Substitution of this into the deterministic form of Eq.~(\ref{e:Duff_sys_x}) gives the dynamics along the slow
manifold as $\dot{x}=x-x^3$.

If, as in Sec.~\ref{sec:gen_theory_DCM}, we let $t=\epsilon\tau$ and denote $\dot{}$ as
$d/dt$ and $^{\prime}$ as $d/d\tau$, then Eqs.~(\ref{e:Duff_sys_x}) and~(\ref{e:Duff_sys_y}) (with $\phi(t)=0$)
are transformed to the following system:
\begin{subequations}
\begin{equation}
\label{e:Duff_trans_x}
x^{\prime}=\epsilon y, 
\end{equation}
\begin{equation}
\label{e:Duff_trans_y}
y^{\prime}=x-x^3-y, 
\end{equation}
\begin{equation}
\label{e:Duff_trans_eps}
\epsilon^{\prime}=0.
\end{equation}
\end{subequations}
Rearrangement of Eqs.~(\ref{e:Duff_trans_x})-(\ref{e:Duff_trans_eps}) leads to
a system described by constant matrices ${\bf \bar{A}}$ and ${\bf B}$ that satisfy the spectral
requirements of Sec.~\ref{sec:gen_theory_DCM}.  Furthermore, since the $x$ and $\epsilon$
variables are associated with the ${\bf \bar{A}}$ matrix (eigenvalues with
zero real parts), and
the $y$ variable is associated with the ${\bf B}$ matrix (eigenvalues with
negative real parts), we know that the
center manifold is given by $y=h(x,\epsilon)$.

The center manifold condition is given by Eq.~(\ref{e:cent_man_cond}), and we
approximate the center manifold [Eq.~(\ref{e:expand})] as follows:
\begin{subequations}
\begin{flalign}
h(x,\epsilon)&=h_0(x)+\epsilon h_1(x)+\epsilon^2 h_2(x)+\mathcal{O}(\epsilon^3)\label{e:h}\\
\nonumber\\
&=c_0 + c_{01}\epsilon +c_{10}x +c_{02}\epsilon^2 + c_{11}x\epsilon +
c_{20}x^2 \nonumber\\
&+ c_{03}\epsilon^3 + c_{12}x\epsilon^2 + c_{21}x^2\epsilon +
c_{30}x^3 + \mathcal{O}(\gamma^4),\label{e:h2}
\end{flalign}
\end{subequations}
where $c_0$, $c_{01}$, $c_{10}$, $c_{02}$, $\ldots$ are unknown coefficients,
and $\gamma=|(x,\epsilon)|$ so that $\gamma$ provides a count of the number of
$x$ and $\epsilon$ factors in any one term.
The center manifold condition for this example
is given by
\begin{equation}
\label{e:Duff_cent_man_cond}
\frac{\partial h(x,\epsilon)}{\partial x}[\epsilon h(x,\epsilon)]=-h(x,\epsilon)+x-x^3.
\end{equation}

By substituting Eq.~(\ref{e:h2}) into Eq.~(\ref{e:Duff_cent_man_cond}) and matching the
different orders to find the coefficients, one finds the following center
manifold equation (expanded to sixth-order):
\begin{subequations}
\begin{flalign}
h(x,\epsilon) = &\,\, x -\epsilon x +2\epsilon^2 x -x^3 -5\epsilon^3 x +
4\epsilon x^3 +14\epsilon^4 x \nonumber\\
&-20\epsilon^2 x^3 -42\epsilon^5 x +104\epsilon^3
x^3 -3\epsilon x^5 +\mathcal{O}(\gamma^7)\label{e:h3}\\
\nonumber\\
=&\, \, x-x^3+\epsilon (-x+4x^3-3x^5) + \epsilon^2 (2x-20x^3)\nonumber\\
& +\epsilon^3
(-5x+104x^3)+\epsilon^4 (14x) +\epsilon^5 (-42x)+\mathcal{O}(\gamma^7).\label{e:h4}
\end{flalign}
\end{subequations}
Note that by letting $\epsilon =0$, one recovers the zero-order approximation,
$h_0(x)$ (the slow manifold).  In addition, since $\epsilon$ is now a state
variable, the first nontrivial correction term to the zero-order approximation
is a quadratic term.

\subsection{Optimal Escape Path and Escape Rate}\label{sec:example_OEP}
Consider the third-order center manifold equation
given by $h(x,\epsilon)=x-x^3-\epsilon x+2\epsilon^2 x$ [see
Eq.~(\ref{e:h3})].  Following Sec.~\ref{sec:gen_theory_OEP}, the dynamics on the
center manifold are given by $x^{\prime}=\epsilon h(x,\epsilon)$
[see Eq.~(\ref{e:dyn})].  Use of the relation $t=\epsilon\tau$ leads to the following
deterministic equation:
\begin{equation}
\label{e:xdot}
\dot{x}=h(x,\epsilon)=x-x^3-\epsilon x +2\epsilon^2 x.
\end{equation}

We ``na{\"i}vely'' add the noise term $\phi(t)$ to the right-hand side of
Eq.~(\ref{e:xdot}) as in Eqs.~(\ref{e:dyn_stoch}) and~(\ref{e:dyn_stoch_1D}) so that
\begin{equation}
\label{e:LangEq}
\dot{x}=x-x^3-\epsilon x +2\epsilon^2 x+\sqrt{2D}\phi(t).
\end{equation}

Equation~(\ref{e:LangEq}) corresponds to a Langevin equation of a particle in an
over-damped quartic potential well with stable states (attractors) located at
$x=x_a=\pm \sqrt{1-\epsilon+2\epsilon^2}$ and an unstable state (saddle)
located at $x=x_s=0$.  

The probability of optimal escape is given by Eqs.~(\ref{e:opt_esc})
and~(\ref{e:R}).  Solution of the Euler-Lagrange equation of motion leads to
the following optimal escape path:
\begin{equation}
\label{e:opt_esc_path}
x_{\rm esc}=\sqrt{\frac{A}{1+3\exp{(2At)}}},
\end{equation} 
where $A=1-\epsilon  +2\epsilon^2$.  Note that $x_{\rm esc}\to \sqrt{A}$ as
$t\to -\infty$ while $x_{\rm esc}\to 0$ as
$t\to \infty$.  This path is a heteroclinic orbit from the basin of attraction
located at
$x=\sqrt{A}$ to the saddle located at $x=0$.

Using Eq.~(\ref{e:opt_esc_path}) along with 
\begin{equation}
\dot{x}_{\rm esc}=x_{\rm esc}-x_{\rm esc}^3-\epsilon x_{\rm esc} +2\epsilon^2
x_{\rm esc}+\sqrt{2D}\phi_{\rm opt}(t),
\end{equation}
we find that the optimal noise is given by the following:
\begin{equation}
\label{e:opt_noise}
\phi_{\rm
  opt}(t)=\frac{1}{\sqrt{2D}}\times\sqrt{\frac{A}{1+3\exp{(2At)}}}\times\left
  [\frac{-6A\exp{(2At)}}{1+3\exp{(2At)}}\right ].
\end{equation}
Since $A$ is a function of $\epsilon$, the shape of $\phi_{\rm opt}(t)$ will
be affected by the value of $\epsilon$.  With $D=0.05$, Fig.~\ref{fig:phi_opt}(a) shows
$\phi_{\rm opt}(t)$ for various values of $\epsilon$.  To obtain a clearer view,
Fig.~\ref{fig:phi_opt}(b) shows a section of Fig.~\ref{fig:phi_opt}(a).  One
can see in Figs.~\ref{fig:phi_opt}(a) and~\ref{fig:phi_opt}(b) that $\epsilon$
has an effect on both the
pulse width and amplitude.  Starting with $\epsilon =0.02$, the pulse
amplitude decreases monotonically and the pulse width increases monotonically
as $\epsilon$ increases to $\epsilon =0.25$ (the value of $\epsilon$ for which
$A$ is minimized).  As $\epsilon$ continues to increase beyond $\epsilon
=0.25$, the pulse amplitude increases monotonically and the pulse width
decreases monotonically. 

Using the theory outlined in Sec.~\ref{sec:gen_theory_OEP}, we now derive an expression for the
escape rate from one of the attractors to the
saddle in order to predict the change in escape rate caused by varying the value of $\epsilon$ and
$D$.  

Since the stochastic (``na{\"i}ve''), third-order center manifold dynamical equation
given by Eq.~(\ref{e:LangEq}) has the form of Eq.~(\ref{e:U_eq}), then the
escape rate from the attractor located at
$x=x_a=\sqrt{1-\epsilon+2\epsilon^2}$ to the saddle located at $x=x_s=0$ can
be found using Eq.~(\ref{e:W}).  The escape rate is given as follows:
\begin{subequations}
\begin{equation}
\label{e:3rd_W}
W(\epsilon
,D)=\frac{\sqrt{|-1+\epsilon-2\epsilon^2|(2-2\epsilon+4\epsilon^2)}}{2\pi}\times \exp{(-\Delta U/D)},
\end{equation}
\begin{equation}
\label{e:3rd_DU}
\Delta U = \left
  |\frac{-1+2\epsilon -5\epsilon^2 +4\epsilon^3 -4\epsilon^4}{4} \right |.
\end{equation}
\end{subequations}
Appendix~\ref{sec:ERSCM} contains similar expressions for the escape rate that are found
using the fourth-order and fifth-order stochastic (``na{\"i}ve'') center
manifold dynamical equations.    

 \subsection{Stochastic Center Manifold and the Normal Form Coordinate Transform}\label{sec:example_SCM}

To more accurately describe the stochastic effects, we will derive the
normal form coordinate transform (and thus the stochastic center manifold) for the singularly perturbed, stochastic Duffing system given by
Eqs.~(\ref{e:Duff_sys_x}) and~(\ref{e:Duff_sys_y}).  As demonstrated previously, use of
the $t=\epsilon \tau$ transformation leads to the following system:\clearpage
\begin{subequations}
\begin{equation}
\label{e:Duff_trans_x_stoch}
x^{\prime}=\epsilon (y +\sqrt{2D}\phi)=\epsilon(y+\sigma\phi),
\end{equation}
\begin{equation}
\label{e:Duff_trans_y_stoch}
y^{\prime}=x-x^3-y,
\end{equation}
\begin{equation}
\label{e:Duff_trans_eps_stoch}
\epsilon^{\prime}=0,
\end{equation}
\end{subequations}  
where $\sigma$ is the
standard deviation of the noise intensity $D=\sigma^2/2$.

The construction of the
normal form is quite tedious and complicated.  However, the result allows one
to determine if there are any noise terms that cause a significant difference between the
average stochastic center manifold (the stochastic center manifold generally
fluctuates about an average location) and the deterministic center manifold.

For this problem, it turns out that the noise terms that could lead to a
difference between the deterministic and average stochastic center manifolds occur at very high order in the
normal form expansion.  Therefore, the correction to the deterministic center
manifold is minimal, and we expect that the deterministic results of
Sec.~\ref{sec:example_DCM} and Sec.~\ref{sec:example_OEP} will agree very well with numerical computations using the
original stochastic system [Eqs.~(\ref{e:Duff_sys_x}) and~(\ref{e:Duff_sys_y})].

We proceed by showing how to use the method
of~\cite{rob08} described in Sec.~\ref{sec:gen_theory_SCM} to construct a normal form coordinate
transform that separates the slow and fast dynamics of
Eqs.~(\ref{e:Duff_trans_x_stoch}) and~(\ref{e:Duff_trans_y_stoch}).  In what follows, we outline the steps involved in the first iteration, while details regarding the higher iterations are provided in the appendices.

\subsubsection{First Iteration}\label{example_SCM_first}

We begin by letting
\begin{subequations}
\begin{equation}
\label{e:xNF}
x\approx X,
\end{equation}
\begin{equation}
\label{e:XdotNF}
X^{\prime}\approx 0,
\end{equation}
\end{subequations}
and by finding a change to the $y$ coordinate (fast process) with the form
\begin{subequations}
\begin{equation}
\label{e:yNF}
y = Y+\eta(\tau,X,Y)+\ldots,
\end{equation}
\begin{equation}
\label{e:YdotNF}
Y^{\prime} = -Y+G(\tau,X,Y)+\ldots,
\end{equation}
\end{subequations}
where $\eta$ and $G$ are small corrections to the coordinate transform and the
corresponding evolution equation.  Substitution of
Eqs.~(\ref{e:xNF})-(\ref{e:YdotNF}) into Eq.~(\ref{e:Duff_trans_y_stoch})
gives the following equation:
\begin{equation}
Y^{\prime}+\frac{\partial \eta}{\partial \tau} + \frac{\partial \eta}{\partial
  X}\frac{\partial X}{\partial \tau} + \frac{\partial \eta}{\partial
  Y}\frac{\partial Y}{\partial \tau} = -Y -\eta +X-X^3.
\end{equation}
Replacing $Y^{\prime}=\partial Y/\partial \tau$ with $-Y+G$ [Eq.~(\ref{e:YdotNF})],
noting that $\partial X/\partial \tau =0$ [Eq. (\ref{e:XdotNF})], and ignoring the
term $\partial\eta/\partial Y\cdot G$ since it is a product of small
corrections leads to the following:
\begin{equation}
\label{e:Getaeqn}
G+\frac{\partial \eta}{\partial \tau} -Y \frac{\partial \eta}{\partial
  Y}+\eta = X-X^3.
\end{equation}

Equation~(\ref{e:Getaeqn}) must now be solved for $G$ and $\eta$.  In order to
keep the evolution equation [Eq.~(\ref{e:YdotNF})] as simple as possible
(principle~(\ref{four}) of Sec.~\ref{sec:gen_theory_SCM}), we let
$G=0$, which means that the coordinate transform [Eq.~(\ref{e:yNF})] is modified by $\eta =X-X^3$.
Therefore, the new approximation of the coordinate transform and its dynamics
are given by
\begin{subequations}
\begin{equation}
\label{e:yNF2_a}
y = Y+X-X^3+\mathcal{O}(\zeta^2),
\end{equation}
\begin{equation}
\label{e:yNF2_b}
Y^{\prime} = -Y+\mathcal{O}(\zeta^2),
\end{equation}
\end{subequations}
where $\zeta =|(X,Y,\epsilon,\sigma)|$ so that $\zeta$ provides a count of the
number of $X$, $Y$, $\epsilon$, and $\sigma$ factors in any one term. 

\subsubsection{Higher Iterations}\label{sec:example_SCM_higher}

The construction of the normal form continues by seeking corrections, $\xi$ and $F$, to the
$x$ coordinate transform and the $X$ evolution using the
updated residual of the $x$ equation [Eq.~(\ref{e:Duff_trans_x_stoch})], and
by seeking corrections, $\eta$ and $G$, to the
$y$ coordinate transform and the $Y$ evolution equation using the
updated residual of the $y$ equation [Eq.~(\ref{e:Duff_trans_y_stoch})].
Details regarding the second iteration can be found in Appendix~\ref{sec:SID}.

The
derivation of $\xi$ and $F$ in the second and fourth iterations along with the derivation
of $\eta$ and $G$ in the third iteration leads to the following updated approximation of the coordinate transforms and
their corresponding evolution equations:
\begin{subequations}
\begin{flalign}
y=&Y+X-X^3+\epsilon\left (-X+4X^3-3X^5\right )\nonumber\\
&+\epsilon\sigma\left (-e^{-\tau}*\phi+3X^2e^{-\tau}*\phi\right )+3\epsilon^2XY^2+\mathcal{O}(\zeta^3),\label{e:yNF3_a}
\end{flalign}
\begin{flalign}
Y^{\prime}=&-Y+\epsilon\left (-Y+3X^2Y\right )+\mathcal{O}(\zeta^3),\label{e:yNF3_b}
\end{flalign}
\end{subequations}
\begin{subequations}
\begin{flalign}
x=&X-\epsilon Y+\epsilon^2\left (Y-3X^2Y\right )\nonumber\\
&+\epsilon^2\sigma\left (e^{-\tau}*\phi-3X^2e^{-\tau}*\phi\right )+\mathcal{O}(\zeta^4),\label{e:xNF4_a}
\end{flalign}
\begin{flalign}
X^{\prime}=&\epsilon\left (X-X^3\right )+\epsilon\sigma\phi+\epsilon^2\left
  (-X+4X^3-3X^5\right )\nonumber\\
&+\epsilon^2\sigma\left (-\phi+3X^2\phi\right )+\mathcal{O}(\zeta^4),\label{e:xNF4_b}
\end{flalign}
\end{subequations}
where
\begin{equation}
e^{-\tau}*\phi = \int\limits_{-\infty}^\tau \exp{\left [-(\tau-s)\right ]}\phi(s) \, ds.
\end{equation}
Details regarding the derivation of Eqs.~(\ref{e:yNF3_a}) and~(\ref{e:yNF3_b})
can be found in Appendix~\ref{sec:TID},
while details of the derivation of Eqs.~(\ref{e:xNF4_a}) and~(\ref{e:xNF4_b})
can be found in Appendix~\ref{sec:FID}.

One can continue this iterative procedure to obtain higher order terms in the
expansions of the coordinate transform and normal form.  For the stochastic
Duffing system under consideration, the fifth and sixth iterations lead to
updated approximations of the $x$ and $y$ coordinate transforms (along with
their associated evolution equations) that are extremely long and
complicated.  These approximations can be found in Appendix~\ref{sec:NFCT}.

In the higher order transform given by Eqs.~(\ref{e:yNF5_a}) -(\ref{e:xNF6_b}),
one can see the appearance of quadratic noise terms.  For example, one can see
terms of the form $e^{-\tau}*(e^{-\tau}*\phi)^2$ in the coordinate transforms [Eqs.~(\ref{e:yNF5_a}) and~(\ref{e:xNF6_a})], and one can see terms of
the form $\phi e^{-\tau}*\phi$ in one of the evolution equations
[Eq.~(\ref{e:xNF6_b})].  This quadratic noise is
important because it leads to the creation of a deterministic drift within the
slow dynamics~\cite{rob08,namlin91}.  Furthermore, the stochastic center
manifold generally undergoes fluctuations about a mean or average location.
This average stochastic center manifold is usually different from the
deterministic center manifold, and it is the quadratic noise process that
generates this difference.

\subsubsection{Comparison with Deterministic Center Manifold and Effect of
  Quadratic Noise}\label{sec:example_SCM_comparison}

Letting $Y=0$ and $\sigma =0$ in Eqs.~(\ref{e:yNF5_a})
and~(\ref{e:xNF6_a}) leads to the following deterministic center manifold equation:
\begin{subequations}
\begin{flalign}
\label{e:dCM_x}
x=&X,\\
\nonumber\\
y=&X-X^3+\epsilon(-X+4x^3-3X^5)+\epsilon^2(2X-20X^3+42X^5)\nonumber\\
&+\epsilon^3(-X+16X^3-66X^5+96X^7-45X^9)+\mathcal{O}(\epsilon^3).\label{e:dCM_y}
\end{flalign}
\end{subequations}
Comparison of Eqs.~(\ref{e:dCM_x}) and~(\ref{e:dCM_y}) with Eq.~(\ref{e:h4})
shows agreement through the $\mathcal{O}(\epsilon^2)$ terms.  There appears to be a discrepancy at
order $\mathcal{O}(\epsilon^3)$.  However, we have checked that this apparent discrepancy is
resolved by expanding the stochastic normal form coordinate transform to even
higher order.  For example, the seventh iteration will yield a $-4\epsilon^3
X$ term in the $y$ coordinate transform.  When added to the existing $-\epsilon^3 X$ term, there is an
agreement with the $-5\epsilon^3 x$ term in Eq.~(\ref{e:h4}).

Letting $Y=0$ in Eqs.~(\ref{e:yNF5_a})
and~(\ref{e:xNF6_a}) leads to the stochastic center manifold equation.  If one
takes the expectation of this
stochastic center manifold equation and uses the following identities \cite{rob08}:
\begin{equation}
E[e^{\pm \tau}*\phi]=e^{\pm\tau}*E[\phi],
\end{equation}
\begin{equation}
E[(e^{\pm \tau}*\phi)^2]=\frac{1}{2},
\end{equation}
then one obtains the following:
\begin{flalign}
\label{e:avg_SCM}
E[y]=&X-X^3+\epsilon(-X+4X^3-3X^5)+\epsilon^2(2X-20X^3+42X^5)\nonumber\\
&+\epsilon^3(-X+16X^3-66X^5+96X^7-45X^9)\nonumber\\
&+\epsilon^4\sigma^2(-3X/2+9X^3-27X^5/2),
\end{flalign}
where the $\mathcal{O}(\epsilon^4\sigma^2)$ terms are associated with the quadratic
noise terms in Eq.~(\ref{e:yNF5_a}).

The average stochastic center manifold equation given by Eq.~(\ref{e:avg_SCM})
can now be used in conjunction with the theory outlined in
Sec.~\ref{sec:gen_theory_OEP} to find an analytical expression for the escape rate.  This
calculation has been performed, but because the noise effects occur
at such high order [$\mathcal{O}(\epsilon^4\sigma^2)$], the correction to the
stochastic (``na{\"i}ve'') result is minimal.  

\subsection{Numerical Computation of Escape Time}\label{sec:example_numerics}

When $\sigma =0$ (no noise), the original, singularly perturbed problem given
by Eqs.~(\ref{e:Duff_sys_x}) and~(\ref{e:Duff_sys_y}), has three equilibrium
points given by $(-1,0)$, $(0,0)$, and
$(1,0)$.  At the initial time, $t=0$, a particle
is randomly placed near the stable,
attracting point $(1,0)$ within a circle of radius $0.1$ centered at $(1,0)$.

Equations~(\ref{e:Duff_sys_x}) and~(\ref{e:Duff_sys_y}) are numerically
integrated using a stochastic fourth-order Runge-Kutta scheme~\cite{rum82,hanpen06} with a constant
time step size, $\delta t$, that depends on the value of $\epsilon$ ($\delta t
= 0.01$ for $\epsilon < 0.1$, while $\delta t = 0.1$ for $\epsilon \geq 0.1$),
and the time needed for the particle to escape from the basin of attraction is
determined.  This escape time is based on either the time it takes the
particle to cross the $x<-0.2$ barrier, which means the particle has escaped across the
unstable saddle, and has entered the second basin of attraction with stable,
attracting point $(-1,0)$, or when the maximum time ($10,000,000$ for $\delta
t =0.01$ and $100,000,000$ for $\delta t = 0.1$) has been reached.

This computation was performed for $10,000$ particles, and the mean escape
time was determined.  In these computations, $\epsilon$ ranged from
$\epsilon=0.02$ to $\epsilon=1.5$, and $\sigma$ ranged from $\sigma=0.26$ to
$\sigma=0.5$.  Figure~\ref{fig:MES_contour} shows a contour plot of the natural log of
the mean escape time plotted for the above range of $\epsilon$ and
$1/D=2/\sigma^2$ values.  

By taking a vertical slice of Fig.~\ref{fig:MES_contour}, one can look at a
plot of the natural log of the mean escape time versus $1/D$ for a fixed
value of $\epsilon$.  Figure~\ref{fig:MES_vs_noise}(a) shows a vertical slice
of Fig.~\ref{fig:MES_contour} taken at $\epsilon =0.1$ along with a line of best
fit through these numerically computed data points.  The slope of the best fit line is $m=0.2505$.
From Eq.~(\ref{e:log-MET}), we see that if one plots the natural log of the
mean escape time vs. $1/D$, then the theoretical slope of this line is given
by $\Delta U$, the depth of the potential well.  If $\epsilon = 0$,
the depth of the potential well corresponding to Eqs.~(\ref{e:Duff_sys_x})
and~(\ref{e:Duff_sys_y}) is $\Delta U = 0.25$, which compares very well with
the numerical result for $\epsilon =0.1$.

Figure~\ref{fig:MES_vs_noise}(b) shows several vertical slices of
Fig.~\ref{fig:MES_contour} taken at various values of
$\epsilon$.  As in Fig.~\ref{fig:MES_vs_noise}(a), there is a line of best fit through the
data points corresponding to each choice of $\epsilon$.  

Numerical computations
have been performed for values of $\epsilon$ as small as $0.02$.  The slopes
of the lines of best fit through the data for these small values of $\epsilon$
are very close to $0.25$.  The data and their lines of best fit are not
shown in Fig.~\ref{fig:MES_vs_noise}(b) since the plots would obscure one
another (and they would all be obscured by the $\epsilon =0.1$ plot). 

One can
see in Fig.~\ref{fig:MES_vs_noise}(b) that as the value of
$\epsilon$ increases (which means that the system moves further and further
away from the over-damped regime), the slope of the line of best fit
decreases.  The slope values are as follows:  for $\epsilon =0.2$, $m=0.2466$;
for $\epsilon =0.3$, $m=0.2398$; for $\epsilon =0.4$, $m=0.2287$; for
$\epsilon =0.5$, $m=0.2165$; and for $\epsilon =1.0$, $m=0.1549$.  When the system is in an
under-damped regime, the slope
of the best fit line no longer agrees with the theoretical value when
$\epsilon = 0$ ($m=0.25$).  However, there still is a nice scaling behavior.
One should note that in general the escape rate computed using the
one-dimensional system found by letting $\epsilon = 0$ may be different than
the escape rate of the associated two-dimensional system with $\epsilon\to
0$~\cite{note4}.  A specific example may be found in the case of extinction
processes~\cite{sbdl09}, where the rate to extinction has no limit when a
singular parameter becomes small.  

As shown in
Sec.~\ref{sec:gen_theory_OEP}, Eqs.~(\ref{e:3rd_W}) and~(\ref{e:3rd_DU}) can be used to find
analytical values of the natural log of
the mean escape time.  By varying the value of $D$ and by fixing the value of
$\epsilon$, we can compare the analytical mean escape time found using
Eqs.~(\ref{e:3rd_W}) and~(\ref{e:3rd_DU}) [or
Eqs.~(\ref{e:4th_W})-(\ref{e:4th_DU}); Eqs.~(\ref{e:5th_W})-(\ref{e:5th_DU})] with the numerically computed mean
escape time shown in
Figs.~\ref{fig:MES_contour} and~\ref{fig:MES_vs_noise}. 

With $\epsilon =0.02$, Fig.~\ref{fig:MES_vs_noise_num_and_CM5}(a) shows a
comparison between the numerically computed mean escape
time (Fig.~{\ref{fig:MES_contour}) and the analytically computed mean escape time found using the fifth-order center manifold
  equation [Eqs.~(\ref{e:5th_W})-(\ref{e:5th_DU})].
  Figure~\ref{fig:MES_vs_noise_num_and_CM5}(a) also contains lines of best fit
  passing through the data points associated with numerical and
  analytical computation. 

One can see from Fig.~\ref{fig:MES_vs_noise_num_and_CM5}(a) that there is good
agreement between the two methods.  The slope of the line of best fit through
the numerical data is $m_{\rm num}=0.2536$, while the slope of the line of best fit
through the analytical data is $m_{\rm ana}=\Delta U =0.2591$.  One also can
see in Fig.~\ref{fig:MES_vs_noise_num_and_CM5}(a) that there is a
slight discrepancy between the analytical and numerical $y$-intercept values.
This is due to the fact that in the numerical computation, the escape time was
based on the time for the particle to cross the barrier and descend partially
into the second basin of attraction, while
the analytical escape time was based on the time for the particle to reach the
top of the barrier.

The two methods continue to agree well as $\epsilon$ is increased to $\epsilon
=0.1$ [Fig.~\ref{fig:MES_vs_noise_num_and_CM5}(b)] and $\epsilon =0.14$
[Fig.~\ref{fig:MES_vs_noise_num_and_CM5}(c)].  Beyond $\epsilon =0.14$, the
analytical result begins to diverge from the numerical result, and this
divergence increases as $\epsilon$ increases.  An example of this divergence
is shown for $\epsilon =0.2$ in Fig.~\ref{fig:MES_vs_noise_num_and_CM5}(d).
The slopes of the lines of best fit are as follows:  for $\epsilon = 0.1$,
$m_{\rm num}=0.2505$ and $m_{\rm ana}= \Delta U =0.2624$; for $\epsilon = 0.14$,
$m_{\rm num}=0.2490$ and $m_{\rm ana}= \Delta U =0.2385$; and for $\epsilon = 0.2$,
$m_{\rm num}=0.2466$ and $m_{\rm ana}= \Delta U =0.1859$.  Note that by linearizing the deterministic form of
Eqs.~(\ref{e:Duff_sys_x})-(\ref{e:Duff_sys_y}) about the stable, attracting point $(1,0)$,
one finds that the critical damping value is $\epsilon_{\rm cr} = 0.125$.
This value agrees well with the value at which the good comparison between the
analytical result and numerical
result begins to break down. 

As stated in Sec.~\ref{sec:example_SCM}, we have computed the analytical escape rate of the
particle using the stochastic center manifold.  Again, the stochastic
correction is minimal, and therefore these analytical mean escape times are
very close in value to those found using the ``na{\"i}ve'' approach.  We do
not show figures comparing this analytical result with the numerical result
since there is no noticeable difference from the plots shown in Figs.~\ref{fig:MES_vs_noise_num_and_CM5}(a)-\ref{fig:MES_vs_noise_num_and_CM5}(d).

\subsection{Numerical Computation of Escape Prehistory}\label{sec:example_prehistory}

For each of the $10,000$ particles that were initially placed in one of the
attracting basins and which later escaped from this basin, across the saddle,
and into the other basin of attraction, we retain $t=200$ worth of the
particle's path prior to escape.  By creating a histogram representing the
probability density, $p_h$, of this escape
prehistory~\cite{dmsss92}, one can see which regions of the phase space are associated with a high or low
probability of particle escape. 

An example of a histogram of escape path prehistory is shown in
Fig.~\ref{fig:hist_sig3_eps1_p10000}(a) for $\epsilon=0.1$ and $\sigma = 0.3$ (so
that $D=\sigma^2 /2 = 0.045$).  The color-bar values of
Fig.~\ref{fig:hist_sig3_eps1_p10000}(a) have been normalized by $10^5$.  The
threshold $0$ value in Fig.~\ref{fig:hist_sig3_eps1_p10000}(a) is actually
about $9000$.  Therefore, any histogram box containing less than $9000$ events
shows up as white on the histogram.  Figure~\ref{fig:hist_sig3_eps1_p10000}(b) is
the same as Fig.~\ref{fig:hist_sig3_eps1_p10000}(a), but with adjusted
color-bar values.  Doing this enables one to obtain a clearer view of the
escape path prehistory along the separatrix and near the saddle.
Figure~\ref{fig:hist_sig3_eps1_p10000}(b) also includes the escape path
prehistory for one particular particle.  To avoid clutter, we have overlaid
only $t=50$ of path prehistory for this one particle.

Figure~\ref{fig:hist_sig3_eps1_p10000_man}(a) shows
Fig.~\ref{fig:hist_sig3_eps1_p10000}(a) overlaid with the graph of the slow
manifold equation, $y=x-x^3$.  Even though this equation is
found by setting $\epsilon = 0$ in the deterministic version of
Eqs.~(\ref{e:Duff_sys_x}) and~(\ref{e:Duff_sys_y}), we see in
Fig.~\ref{fig:hist_sig3_eps1_p10000_man}(a) that the slow manifold lies very
close to the region associated with the highest probability of escape.

Similarly, Fig.~\ref{fig:hist_sig3_eps1_p10000_man}(b) shows
Fig.~\ref{fig:hist_sig3_eps1_p10000}(a) overlaid with the graphs of the
third-order, fourth-order,
and fifth-order center manifold
equations.  Each of these equations may be found by including terms of the
appropriate order from Eq.~(\ref{e:h3}).  As in Fig.~\ref{fig:hist_sig3_eps1_p10000}(a), the color-bar values of
Figs.~\ref{fig:hist_sig3_eps1_p10000_man}(a)
and~\ref{fig:hist_sig3_eps1_p10000_man}(b) have been normalized by $10^5$.

One can see from Fig.~\ref{fig:hist_sig3_eps1_p10000_man}(b) that the
third-order and fourth-order center manifolds essentially bound the entire
region of escape path prehistory, while the fifth-order manifold lies along
the region of highest probability of escape.  Although it is not shown, it
should be noted that the optimal escape
path [Eq.~(\ref{e:opt_esc_path})] associated with the
third-order center manifold is a heteroclinic orbit from $x=\sqrt{1-\epsilon
  +2\epsilon^2}$ to $x=0$ that lies directly on top of a section of the third-order center
manifold [solid, green line in
Fig.~\ref{fig:hist_sig3_eps1_p10000_man}(b)].  Similarly, the optimal escape
paths associated with higher order center manifolds lie directly on top of a
section of the corresponding center manifold.

Additionally, one could overlay the histogram of escape path prehistory with
the average stochastic center manifold given by Eq.~(\ref{e:avg_SCM}).
However, since the stochastic correction appears at order
$\mathcal{O}(\epsilon^4\sigma^2)$, there is no noticeable difference from the
manifolds shown in Figs.~\ref{fig:hist_sig3_eps1_p10000_man}(a)
and~\ref{fig:hist_sig3_eps1_p10000_man}(b).  Therefore, plots of the average
stochastic manifold are not shown.

\section{Conclusions}\label{sec:conc}

A general procedure consisting of elements taken from deterministic and stochastic manifold
theory and large fluctuation theory is developed and used to understand the
underlying structure of a stochastic dynamical system with two well-separated
time scales. 

As a first step towards this goal, we have applied the procedure to a generic 2D, singularly
perturbed, damped, Duffing oscillator system with additive, Gaussian noise.
The deterministic center manifold equation is found by neglecting the
stochastic terms.  By ``na{\"i}vely'' adding a noise term to the equation that
describes the dynamics on the center manifold, the path integral formalism of
large fluctuation theory can be used to analytically compute the optimal
escape path of the particle from one basin of attraction to another basin of
attraction as well as the particle's escape rate.  Comparison of the
analytical result with numerical computations shows excellent agreement if the
system is in the over-damped regime.  There is a shift due to 
the pre-factor of the distribution, but the exponent is quite accurate.  While
it is possible to correctly determine the
pre-factor using spectral methods, the technique works only
for 1D problems~\cite{note3}.  When the system enters the under-damped
regime, we must consider the 2D topology rather than the 1D manifold topology
that is associated with the over-damped regime~\cite{dyscsh05}.  We are
currently performing the analysis associated with the under-damped regime, and this
will be presented elsewhere.

Additionally, we used numerical computation to create a histogram of the
escape path prehistory distribution.  The histogram enables one to
see regions of the phase space that are associated with high and low probability
of particle escape from the basin of attraction.  By overlaying the histogram
with the deterministic slow manifold, we see that this manifold lies close to
the region associated
with the highest probability of escape.  By comparing the histogram
with deterministic center manifolds of various orders, we see that the third-order and fourth-order
center manifolds essentially bound the entire region of escape, while the
fifth-order center manifold lies very close to the region associated
with the highest probability of escape.  Therefore one can use these manifolds
to accurately describe the location of escape regions.  The example considered here is that of a quartic potential which is a
  generic potential for a pitchfork bifurcation.  Therefore, for systems whose
  center manifolds yield pitchfork dynamics, we expect similar results to hold
  when additive noise is considered.

Knowledge of the location of the regions of high and low probability of escape
(whether from numerical or analytical results) is extremely useful if one
wishes to optimize the amount of time spent in a particular region of phase space.  For example, one might wish to keep an
autonomous glider in
some basin of attraction.  Instead of constantly actuating the glider controls
to keep the glider stationed at a particular location, one can station the
glider initially in a region of low probability of escape.  Then, if the glider enters a region of high probability
of escape, one actuates the controls to move the glider back to a region with
low probability of escape and the controls are switched off. This is 
similar to the stochastic control used in~\cite{Schwartz2004,Dykman2000}. 

As an example, consider a particle whose dynamics are described by Eqs.~(\ref{e:Duff_sys_x})
and~(\ref{e:Duff_sys_y}).  Figure~\ref{fig:x_vs_t}(a) shows the location of
the particle ($x$ coordinate) as a function of time $t$.  One can see in
Fig.~\ref{fig:x_vs_t}(a) that the
particle moves around the basin of attraction until eventually the stochastic
effects cause the particle to escape from the basin at $t\approx 1290$.  

By
implementing a simple control that pushes the particle into a region of phase
space that has a low probability of escape whenever the particle enters a
region of high probability of escape, it is possible to keep the particle in
the basin of attraction for a much longer amount of time.
Figure~\ref{fig:x_vs_t}(b) shows the location of
the particle ($x$ coordinate) as a function of time $t$ with the application
of control.  One can see in Fig.~\ref{fig:x_vs_t}(b) that control allows one
to keep the particle in the basin of attraction until $t=10,000$, when the
simulation was stopped.  Preliminary results show that the application of control increases the mean time to
escape from the basin and decreases the total number of particles which escape
in a given period of time.  A complete analysis of the application of control
will appear elsewhere. 

Even though we showed that analytical results using a ``na{\"i}ve'' approach
agree very well with the numerical
results in the over-damped regime, to more accurately describe the
stochastic effects, we derived the normal form coordinate transform.  The
normal form enabled us to find the stochastic center manifold.  However, the
stochastic effects appeared in the manifold equation at high order, so this
correction has a minimal effect.  Therefore, for this problem at least, one
can use the simpler and much less time consuming ``na{\"i}ve'' approach.  

It
should be noted that there are systems where one should not rely on the
``na{\"i}ve'' approach.  For example, in a
Susceptible-Exposed-Infected-Recovered (SEIR) epidemiological
model, there are terms at low order in
the normal form transform which cause a significant difference between the
average stochastic center manifold and the deterministic manifold~\cite{fobisc08pre}.  Therefore,
when working with the SEIR model, one must use the stochastic normal form
coordinate transform approach to obtain the correct projection of
the noise onto the center manifold.

Figure~\ref{fig:SEIR}(a) compares the fraction of the population that is
infected with a disease, $I$, computed using the complete, stochastic system
of equations of the SEIR model with the time series of $I$ computed using the
reduced system of equations of the SEIR model that is based on the
deterministic center manifold with a ``na{\"i}ve'' replacement of the noise
terms.  One can see that the solution computed using the reduced system
incorrectly predicts the time and amplitude of the initial outbreak and quickly becomes
out of phase with the solution of the complete system.  Although not shown,
the poor agreement, in phase and amplitude, between the two solutions
continues for long periods of time. 

On the other hand, Fig.~\ref{fig:SEIR}(b) compares the time series of $I$ computed using the complete, stochastic system
of equations of the SEIR model with the time series of $I$ computed using the
reduced system of equations of the SEIR model that is found using the
stochastic normal form coordinate transform.  There is excellent agreement
between the two solutions.  The initial outbreak is successfully captured by
the reduced system, and the reduced system continues to accurately predict the
phase and amplitude of outbreaks over long periods of time. 

The previous general analysis has been performed only for the 2D singularly
perturbed Duffing system.  Of interest is the application of the theory to
more realistic stochastic dynamical systems.  Beyond that, we plan to apply the theory to fully 3D systems as well as
to actual oceanographic data.

\section*{Acknowledgments}
We warmly thank M. Dykman for his insight and for introducing us to the optimal
path theory.  We also thank A.J. Roberts for his initial reading of the
manuscript.  The authors benefited from the comments and suggestions of
anonymous reviewers.  We gratefully acknowledge support from the Office of
Naval Research, the
Army Research Office, and the Air Force Office of Scientific Research.  E.F. is supported by a National Research Council Research Associateship.

\begin{appendix}

\section{Escape Rate Using Stochastic (``Na{\"i}ve'') Center Manifolds}\label{sec:ERSCM}
Using the fourth-order stochastic (``na{\"i}ve'') center manifold dynamical equation given by
\begin{equation}
\label{e:4th_LangEq}
\dot{x}=x-x^3-\epsilon x +2\epsilon^2 x -5\epsilon^3 x+4\epsilon x^3+\sqrt{2D}\phi(t),
\end{equation}
one finds that the escape rate from the attractor located at
\begin{equation}
x=x_a=\sqrt{(1-\epsilon+2\epsilon^2-5\epsilon^3)/(1-4\epsilon)}
\nonumber
\end{equation}
to the
saddle located at $x=x_s=0$
is given as follows:
\begin{subequations}
\begin{equation}
\label{e:4th_W}
W(\epsilon ,D)=\frac{\sqrt{|U^{\prime\prime}(0)|U^{\prime\prime}(x_a)}}{2\pi}\exp{(-\Delta U/D)},
\end{equation}
where 
\begin{equation}
U^{\prime\prime}(0)=-1+\epsilon-2\epsilon^2+5\epsilon^3,
\end{equation}
\begin{equation}
U^{\prime\prime}(x_a)=2-2\epsilon+4\epsilon^2-10\epsilon^3,
\end{equation}
\begin{flalign} 
\label{e:4th_DU}
\Delta U =
    \left | \left (\right. \right.&\left.-1 + 6\epsilon -13\epsilon^2 +34\epsilon^3 -70\epsilon^4
     +76\epsilon^5 \right. \nonumber\\
   & \left.\left. -105\epsilon^6 +100\epsilon^7 \right )/ \left [4
      \left (1-8\epsilon+16\epsilon^2  \right ) \right ] \right|.
\end{flalign}
\end{subequations}

Using the fifth-order stochastic (``na{\"i}ve'') center manifold dynamical equation given by
\begin{equation}
\label{e:5th_LangEq}
\dot{x}=x-x^3-\epsilon x +2\epsilon^2 x-5\epsilon^3 x+4\epsilon
x^3+14\epsilon^4 x-20\epsilon^2 x^3+\sqrt{2D}\phi(t),
\end{equation}
one finds that the escape rate from the attractor located at
\begin{equation}
x=x_a=\sqrt{(1-\epsilon+2\epsilon^2-5\epsilon^3+14\epsilon^4)/(1-4\epsilon
+20\epsilon^2)} \nonumber
\end{equation}
to the saddle located at $x=x_s=0$
is given as follows:
\begin{subequations}
\begin{equation}
\label{e:5th_W}
W(\epsilon ,D)=\frac{\sqrt{|U^{\prime\prime}(0)|U^{\prime\prime}(x_a)}}{2\pi}\exp{(-\Delta U/D)},
\end{equation}
where 
\begin{equation}
U^{\prime\prime}(0)=-1+\epsilon-2\epsilon^2+5\epsilon^3-14\epsilon^4,
\end{equation}
\begin{equation}
U^{\prime\prime}(x_a)=2-2\epsilon+4\epsilon^2-10\epsilon^3+42\epsilon^4,
\end{equation}
\begin{flalign}
\label{e:5th_DU}
\Delta U = \left
  |\left ( \right.\right.&\left.1-2\epsilon+5\epsilon^2-14\epsilon^3+42\epsilon^4-48\epsilon^5 \right.\left. +81\epsilon^6-140\epsilon^7+196\epsilon^8\right )\nonumber\\
&/\left.\left[
    4\left (-1+4\epsilon-20\epsilon^2\right )\right ] \right |.
\end{flalign}
\end{subequations}

\section{ Second Iteration Details}\label{sec:SID}
For this second iteration, we seek a correction to the $x$ coordinate (slow process)
with the form 
\begin{subequations}
\begin{equation}
\label{e:xNF2}
x = X+\xi(\tau,X,Y)+\ldots, 
\end{equation}
\begin{equation}
\label{e:XdotNF2}
X^{\prime} = F(\tau,X,Y)+\ldots,
\end{equation}
\end{subequations}
where $\xi$ and $F$ are small corrections.  Substitution of
Eqs.~(\ref{e:yNF2_a})-(\ref{e:XdotNF2}) into
Eq. (\ref{e:Duff_trans_x_stoch}) leads to
\begin{equation}
X^{\prime}+\frac{\partial \xi}{\partial \tau} + \frac{\partial \xi}{\partial
  X}\frac{\partial X}{\partial \tau} + \frac{\partial \xi}{\partial
  Y}\frac{\partial Y}{\partial \tau} = \epsilon(Y+X-X^3)+\epsilon\sigma\phi.
\end{equation}
Replacing $X^{\prime}=\partial X/\partial \tau$ with $F$ [Eq.~(\ref{e:XdotNF2})],
replacing $\partial Y/\partial \tau$ with $-Y$ [Eq.~(\ref{e:yNF2_b})], and
ignoring the term $\partial\xi/\partial X\cdot F$ since it is a product of
small corrections gives the following equation:
\begin{equation}
\label{e:Fxieqn}
F+\frac{\partial \xi}{\partial \tau} -Y\frac{\partial \xi}{\partial
  Y} = \epsilon(Y+X-X^3)+\epsilon\sigma\phi.
\end{equation}

Equation~(\ref{e:Fxieqn}) must now be solved for $F$ and $\xi$.  As in the
first step, we employ principle~(\ref{four}) and keep the
evolution equation [Eq.~(\ref{e:XdotNF2})] as simple as possible.  However, since the terms
$\epsilon (X-X^3)$ located on the right-hand side of Eq.~(\ref{e:Fxieqn}) do not contain $\tau$ or
$Y$, these terms must be included in F.  Therefore, one piece of $F$ will be
$F=\epsilon (X-X^3)$.  

The remaining deterministic term on the right-hand side
of Eq.~(\ref{e:Fxieqn}) contains $Y$.  This term can therefore be integrated
into $\xi$.  The equation to be solved is 
\begin{equation}
\label{e:Eq_2_a}
-Y \frac{\partial\xi}{\partial Y}=\epsilon Y, 
\end{equation}
whose solution is given as $\xi =-\epsilon Y$.

To abide by principle~(\ref{four}), we would like to integrate the
stochastic piece on the right-hand side of Eq.~(\ref{e:Fxieqn}) into $\xi$, by
solving the equation
\begin{equation}
\label{e:Eq_2_b}
\partial \xi/\partial \tau = \epsilon\sigma\phi.  
\end{equation}
However, the solution of Eq.~(\ref{e:Eq_2_b}) is given by
\begin{equation}
\xi=\epsilon\sigma\int\phi \, d\tau,
\end{equation}
which has secular growth like a Wiener
process.  Since this would violate principle~(\ref{one}), 
we must let
$F=\epsilon\sigma\phi$.

Putting the three pieces together yields $\xi=-\epsilon Y$ and
$F=\epsilon(X-X^3)+\epsilon\sigma\phi$.  Therefore, the new approximation of the coordinate transform and its dynamics
are given by
\begin{subequations}
\begin{equation}
\label{e:xNF3_a}
x=X-\epsilon Y+\mathcal{O}(\zeta^3),
\end{equation}
\begin{equation}
\label{e:xNF3_b}
X^{\prime}=\epsilon(X-X^3)+\epsilon\sigma\phi+\mathcal{O}(\zeta^3).
\end{equation}
\end{subequations}

\section{ Third Iteration Details}\label{sec:TID}
To find the corrections, $\eta$ and $G$, we substitute Eqs.~(\ref{e:yNF}) and~(\ref{e:YdotNF}) into
Eq.~(\ref{e:Duff_trans_y_stoch}), which leads to the following equation:
\begin{equation}
\label{e:Getaeqn_2}
G+\frac{\partial \eta}{\partial \tau} + \frac{\partial \eta}{\partial
  X}\frac{\partial X}{\partial \tau} -Y \frac{\partial \eta}{\partial
  Y} +\eta =x-x^3.
\end{equation}
Substitution of the specific form of $x$ given by Eq.~(\ref{e:xNF3_a}), the
specific form of $X^{\prime}=\partial X/\partial \tau$ given by
Eq.~(\ref{e:xNF3_b}), and substitution of $\eta =X-X^3$ and $G=0$ [see
Eqs.~(\ref{e:yNF2_a}) and~(\ref{e:yNF2_b})] into Eq.~(\ref{e:Getaeqn_2}) gives
one the following evolution equation driven by the updated residual of Eq.~(\ref{e:Duff_trans_y_stoch}):
\begin{flalign}
\label{e:Getaeqn3}
G+\frac{\partial \eta}{\partial \tau} -Y \frac{\partial \eta}{\partial
  Y}+\eta =& \epsilon\left (-X-Y+4X^3-3X^5+3X^2Y\right )\nonumber\\
& + \epsilon\sigma\left (-\phi+3X^2\phi\right ) +\epsilon^2\left (-3XY^2\right
)+\epsilon^3Y^3.
\end{flalign}

We first consider the deterministic terms on the right-hand side of
Eq.~(\ref{e:Getaeqn3}).  Principle~(\ref{four}) is employed to keep the evolution equation
[Eq.~(\ref{e:YdotNF})] as simple as possible, and since the terms $-\epsilon
X$, $4\epsilon X^3$, $-3\epsilon X^5$ are not functions of $\tau$ or $Y$, we
let $\eta =\epsilon (-X+4X^3-3X^5 )$.  Consideration of the term $-3\epsilon^2
XY^2$ leads one to solve the following equation:
\begin{equation}
\label{e:Eq3_a}
\eta-Y\frac{\partial \eta}{\partial Y} = -3\epsilon^2 XY^2.
\end{equation}
The solution of Eq.~(\ref{e:Eq3_a}) is $\eta =3\epsilon^2 XY^2$.

The last deterministic terms on the right-hand side of Eq.~(\ref{e:Getaeqn3})
are $-\epsilon Y$ and $3\epsilon X^2 Y$.  Since these two terms can't be
integrated into $\eta$, they are included in $G$.  The
higher order term $\epsilon^3 Y^3$ will be ignored until a later iteration.

We now consider the stochastic terms on the right-hand side of
Eq.~(\ref{e:Getaeqn3}).  The equation to be solved is
\begin{equation}
\frac{\partial\eta}{\partial \tau} + \eta
=-\epsilon\sigma\phi +3\epsilon\sigma X^2\phi,
\end{equation}
whose solution is given by
\begin{equation}
\label{e:convdef}
\eta =
-\epsilon\sigma\int\limits_{-\infty}^\tau \exp{\left [-(\tau-s)\right ]}\phi(s) \, ds + 3\epsilon\sigma X^2 \int\limits_{-\infty}^\tau \exp{\left [-(\tau-s)\right ]}\phi(s)
\, ds.
\end{equation}
If we define
\begin{equation}
\int\limits_{-\infty}^\tau \exp{\left [-(\tau-s)\right ]}\phi(s) \, ds = e^{-\tau}*\phi,
\end{equation}
then Eq.~(\ref{e:convdef}) can be written as follows:
\begin{equation}
\eta =-\epsilon\sigma e^{-\tau}*\phi + 3\epsilon\sigma X^2e^{-\tau}*\phi.
\end{equation}

Putting all of the $\eta$ and $G$ pieces from this third iteration together
leads to the updated approximation of the coordinate transform and
its evolution equation given by Eqs.~(\ref{e:yNF3_a}) and~(\ref{e:yNF3_b}).

\section{Fourth Iteration Details}\label{sec:FID}
To find the corrections, $\xi$ and $F$, we substitute Eqs.~(\ref{e:xNF2}) and~(\ref{e:XdotNF2}) into
Eq.~(\ref{e:Duff_trans_x_stoch}), which leads to the following equation:
\begin{equation}
\label{e:Fxi_xNF4_2}
F+\frac{\partial \xi}{\partial \tau} + \frac{\partial \xi}{\partial
  X}\frac{\partial X}{\partial \tau} + \frac{\partial \xi}{\partial
  Y}\frac{\partial Y}{\partial \tau} = \epsilon y+\epsilon\sigma\phi.
\end{equation}
Substitution of the specific forms of $y$ [Eq.~(\ref{e:yNF3_a})],
$Y^{\prime}=\partial Y/\partial\tau$ [Eq.~(\ref{e:yNF3_b})], $\xi$ [Eq.~(\ref{e:xNF3_a})], and $F$ [Eq.~(\ref{e:xNF3_b})] into Eq.~(\ref{e:Fxi_xNF4_2}) leads to
the following evolution equation driven by the updated residual of Eq.~(\ref{e:Duff_trans_x_stoch}):
\begin{flalign}
\label{e:Fxieqn4}
F+\frac{\partial \xi}{\partial \tau} -Y\frac{\partial \xi}{\partial Y}
=& \epsilon^2\left (-X+4X^3-3X^5-Y+3X^2Y\right )\nonumber\\
&+\epsilon^2\sigma\left (-e^{-\tau}*\phi+3X^2e^{-\tau}*\phi\right )+3\epsilon^3XY^2.
\end{flalign}
It is straightforward (and similar to what has been done in the previous
iterations) to integrate the deterministic terms on the right-hand side of
Eq.~(\ref{e:Fxieqn4}) into $F$ and $\xi$.  

Consideration of the stochastic terms on the right-hand side of
Eq.~(\ref{e:Fxieqn4}) means that we must solve the following equation:
\begin{equation}
\label{e:It4st}
F+\frac{\partial \xi}{\partial \tau}=\epsilon^2\sigma\left
  (-e^{-\tau}*\phi+3X^2e^{-\tau}*\phi\right ).
\end{equation}
As in the second iteration, we can't integrate $\phi$ into $\xi$, since this would
generate secular growth in violation of principle~(\ref{one}).  Employing principle~(\ref{five}) to avoid a fast time
convolution (memory integral) in the slow evolution $F$ leads us to perform an integration by parts on each of the terms on the right-hand side of
Eq.~(\ref{e:It4st}) so that
\begin{equation}
\label{e:It4st2}
F+\frac{\partial \xi}{\partial \tau}=\epsilon^2\sigma\left
  [-\phi+e^{-\tau}*\phi^{\prime}+3X^2\left (\phi-e^{-\tau}*\phi^{\prime}\right
  )\right ].
\end{equation}
It is clear that we now have $F=\epsilon^2\sigma (-\phi+3X^2\phi )$ and $\xi=\epsilon^2\sigma (e^{-\tau}*\phi-3X^2e^{-\tau}*\phi )$.  

Putting all of the $\xi$ and $F$ pieces from this fourth iteration together
leads to the updated approximation of the coordinate transform and
its evolution equation given by Eqs.~(\ref{e:xNF4_a}) and~(\ref{e:xNF4_b}).

\section{Normal Form Coordinate Transform}\label{sec:NFCT}
\begin{subequations}
\begin{flalign}
y=&Y+X-X^3+\epsilon\left (-X+4X^3-3X^5\right )+\epsilon\sigma\left (-e^{-\tau}*\phi+3X^2e^{-\tau}*\phi \right )\nonumber\\
&+\epsilon^2\left (2X-20X^3 +42X^5+3XY^2\right )\nonumber\\
&+\epsilon^2\sigma
\left (2e^{-\tau}*\phi +e^{-\tau}*e^{-\tau}*\phi\right.-18X^2e^{-\tau}*\phi
-12X^2 e^{-\tau}*e^{-\tau}*\phi\nonumber\\
&\hspace{1.3cm}+24X^4e^{-\tau}*\phi\left. +15X^4e^{-\tau}*e^{-\tau}*\phi\right
)\nonumber\\
&+\epsilon^3 \left (-X+16X^3-66X^5+96X^7-45X^9 \right.\left. -9XY^2
  -Y^3/2+33X^3Y^2 \right ) \nonumber\\
&+\epsilon^3\sigma
\left (-e^{-\tau}*\phi +18X^3Ye^{-\tau}*\phi \right.-6XYe^{-\tau}*\phi +15X^2e^{-\tau}*\phi\nonumber\\
&\hspace{1.3cm}+45X^6e^{-\tau}*\phi -51X^4e^{-\tau}*\phi +18X^6e^{-\tau}*e^{-\tau}*\phi\nonumber\\
&\hspace{1.3cm}-24X^4e^{-\tau}*e^{-\tau}*\phi +6X^2e^{-\tau}*e^{-\tau}*\phi \left. +3Y^2e^{+\tau}*\phi\right )\nonumber\\
&+\epsilon^4 \left (18X^5Y^2-6X^3Y^2+3Y^3/2\right.\left.-9Y^3X^2/2 \right )\nonumber\\
&+\epsilon^4\sigma \left (6XYe^{-\tau}*\phi +54X^5Ye^{-\tau}*\phi \right. -36X^3Ye^{-\tau}*\phi +9X^2Y^2e^{+\tau}*\phi \nonumber\\
&\hspace{1.3cm}-3Y^2e^{+\tau}*\phi +3Y^2e^{+\tau}*e^{-\tau}*\phi \left. -9X^2Y^2e^{+\tau}*e^{-\tau}*\phi\right )\nonumber\\
&+\epsilon^4\sigma^2\left (-3Xe^{-\tau}*\left (e^{-\tau}*\phi\right )^2 \right. -27X^5e^{-\tau}*\left (e^{-\tau}*\phi\right )^2 \nonumber\\
&\hspace{1.3cm}\left. +18X^3e^{-\tau}*\left (e^{-\tau}*\phi\right )^2\right )+\mathcal{O}(\zeta^4),
\label{e:yNF5_a}
\end{flalign}
\begin{flalign}
Y^{\prime}=&-Y+\epsilon\left (-Y+3X^2Y\right )+\epsilon^2 \left
  (Y-6X^2Y+9X^4Y\right )\nonumber\\
&+\epsilon^3\sigma\left (6XY\phi-18X^3Y\phi\right )\nonumber\\
&+\epsilon^4\sigma\left
  (-6XY\phi-54X^5Y\phi\right.\left.+36X^3Y\phi\right )+\mathcal{O}(\zeta^4),
\label{e:yNF5_b}
\end{flalign}
\begin{flalign}
x=&X-\epsilon
Y+\epsilon^2\left (Y-3X^2Y\right )+\epsilon^2\sigma\left
  (e^{-\tau}*\phi-3X^2e^{-\tau}*\phi \right )\nonumber\\
&+\epsilon^3\left (6X^2Y-12X^4Y-2Y-3XY^2/2\right )\nonumber\\
&+\epsilon^4\left (-9X^6Y+3X^4Y-3X^2Y+Y+Y^3/6\right.\left. +9XY^2/2-33X^3Y^2/2\right )\nonumber\\
&+\epsilon^5\left (-Y^3/2+3X^3Y^2+3X^2Y^3/2\right.\left.-9X^5Y^2\right )\nonumber\\
&+\epsilon^3\sigma\left (-3e^{-\tau}*\phi
-33X^4e^{-\tau}*\phi\right. +24X^2e^{-\tau}*\phi  -15X^4e^{-\tau}*e^{-\tau}*\phi\nonumber\\
&\hspace{1.3cm}-6XYe^{+\tau}*\phi -e^{-\tau}*e^{-\tau}*\phi\left.-12X^2e^{-\tau}*e^{-\tau}*\phi\right
)\nonumber\\
&+\epsilon^4\sigma\left (e^{-\tau}*\phi
-15X^2e^{-\tau}*\phi \right.-6X^2e^{-\tau}*e^{-\tau}*\phi +51X^4e^{-\tau}*\phi
\nonumber\\
&\hspace{1.3cm}-45X^6e^{-\tau}*\phi+24X^4e^{-\tau}*e^{-\tau}*\phi -18X^6e^{-\tau}*e^{-\tau}*\phi
\nonumber\\
&\hspace{1.3cm}+3XYe^{+\tau}*\phi
+3XYe^{-\tau}*\phi -9X^3Ye^{+\tau}*\phi -9X^3Ye^{-\tau}*\phi\nonumber\\
&\hspace{1.3cm}\left. -3Y^2e^{+2\tau}*e^{+\tau}*\phi\right
)\nonumber\\
&+\epsilon^5\sigma\left (-54X^3Ye^{+\tau}*\phi +9XYe^{+\tau}*\phi \right.+81x^5Ye^{+\tau}*\phi -3XYe^{-\tau}*\phi \nonumber\\
&\hspace{1.3cm}-27X^5Ye^{-\tau}*\phi +18X^3Ye^{-\tau}*\phi -9X^2Y^2e^{+2\tau}*e^{+\tau}*\phi \nonumber\\
&\hspace{1.3cm}+3\left (Y^2e^{+2\tau}*e^{+\tau}*\phi\right ) /2 \left. -3\left (Y^2e^{+2\tau}*e^{-\tau}*\phi\right ) /2\right ) \nonumber\\
&+\epsilon^6\sigma\left (-6XYe^{+\tau}*\phi +54X^3Ye^{+\tau}*\phi\right. -162X^5Ye^{+\tau}*\phi\left.+162X^7Ye^{+\tau}*\phi\right )\nonumber\end{flalign}\\
\begin{flalign}
&+\epsilon^5\sigma^2\left (3X\left (e^{-\tau}*\phi\right )^2/2 -9X^3\left (e^{-\tau}*\phi\right)^2\right.+27X^5\left (e^{-\tau}*\phi\right )^2/2 &&\nonumber\\
&\hspace{1.3cm}+3Xe^{-\tau}*\left
  (e^{-\tau}*\phi\right )^2 -18X^3e^{-\tau}*\left (e^{-\tau}*\phi\right )^2\nonumber\\
&\hspace{1.2cm}\left. +27X^5e^{-\tau}*\left (e^{-\tau}*\phi\right )^2\right )+\mathcal{O}(\zeta^4),\label{e:xNF6_a}
\end{flalign}
\begin{flalign}
X^{\prime}=&\epsilon\left (X-X^3\right )+\epsilon\sigma\phi +\epsilon^2\left
  (-X+4X^3-3X^5\right )+\epsilon^2\sigma\left (-\phi+3X^2\phi\right )\nonumber\\
&+\epsilon^3\left (2X-20X^3+42X^5\right )\nonumber\\
&+\epsilon^4\left (-X+16X^3-66X^5\right.\left.96X^7-45X^9\right)\nonumber\\
&+\epsilon^3\sigma\left (3\phi -24X^2\phi
+33X^4\phi\right )+\epsilon^4\sigma\left (-\phi
+15X^2\phi -51X^4\phi +45X^6\phi\right )\nonumber\\
&+\epsilon^4\sigma^2\left (-6X\phi e^{-\tau}*\phi +18X^3\phi
e^{-\tau}*\phi\right ) \nonumber\\
&+\epsilon^5\sigma^2\left (-9X\phi e^{-\tau}*\phi +54X^3\phi
  e^{-\tau}*\phi\right. 
\left. -81X^5\phi e^{-\tau}*\phi\right )+\mathcal{O}(\zeta^4),\label{e:xNF6_b}
 \end{flalign}
\end{subequations}
where
\begin{equation}
e^{+\tau}*\phi = \int\limits_\tau^{+\infty} \exp{\left [(\tau-s)\right ]}\phi(s) \, ds.
\end{equation}
\end{appendix}


\clearpage
\begin{figure}[h!]
\begin{center}
\begin{minipage}{0.98\linewidth}
\includegraphics[width=12.5cm]{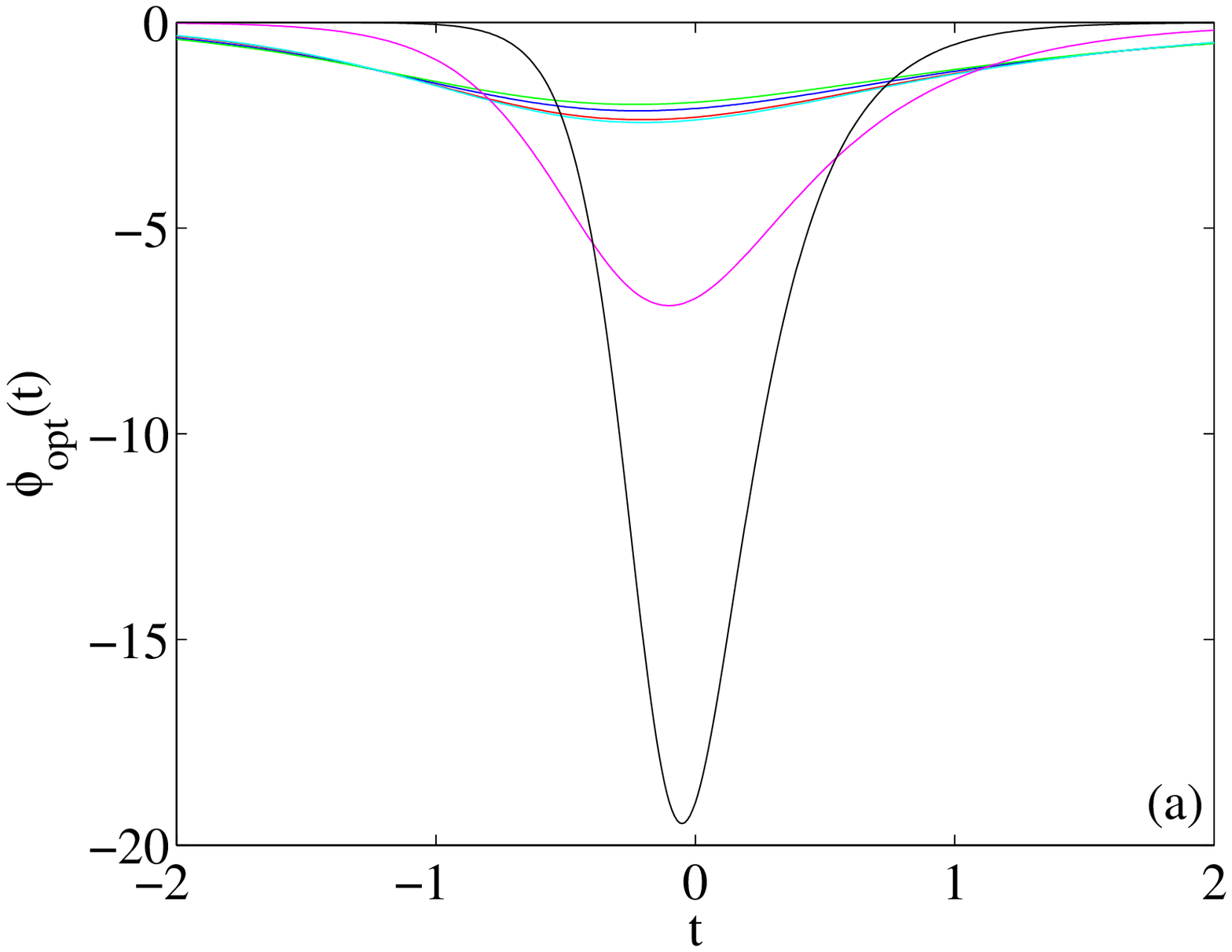}
\end{minipage}\\
\begin{minipage}{0.98\linewidth}
\includegraphics[width=12.5cm]{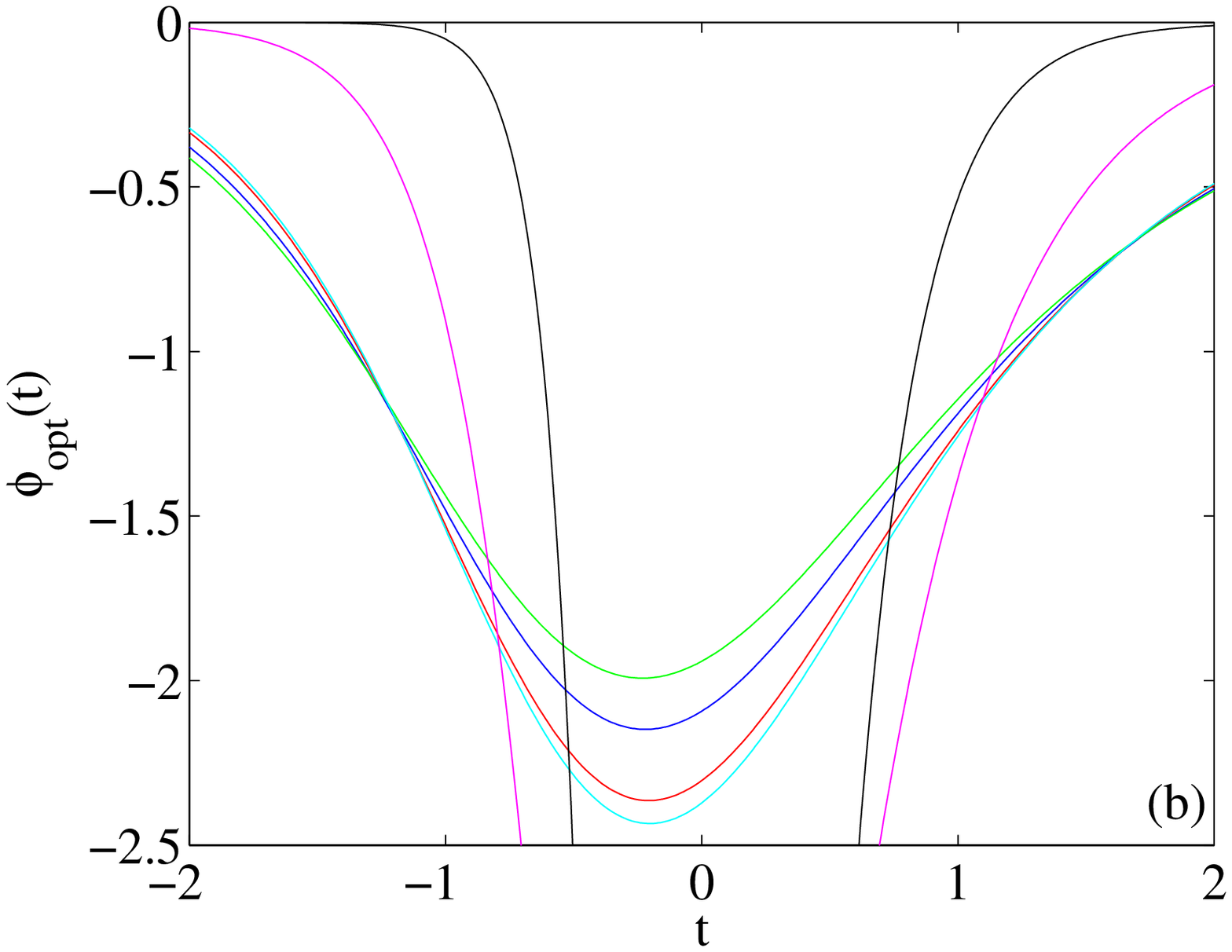}
\end{minipage}
\caption{\label{fig:phi_opt}(a) $\phi_{\rm opt}(t)$ with $D=0.05$ for $\epsilon =0.02$ (red), $\epsilon =0.1$ (blue),
$\epsilon =0.25$ (green), $\epsilon =0.5$ (cyan), $\epsilon =1.0$ (magenta), and
$\epsilon =1.5$ (black).  (b) A close-up view of a section of
Fig.~\ref{fig:phi_opt}(a).}
\end{center}
\end{figure}
\clearpage
\begin{figure}[h!]
\begin{center}
\includegraphics[width=12.5cm]{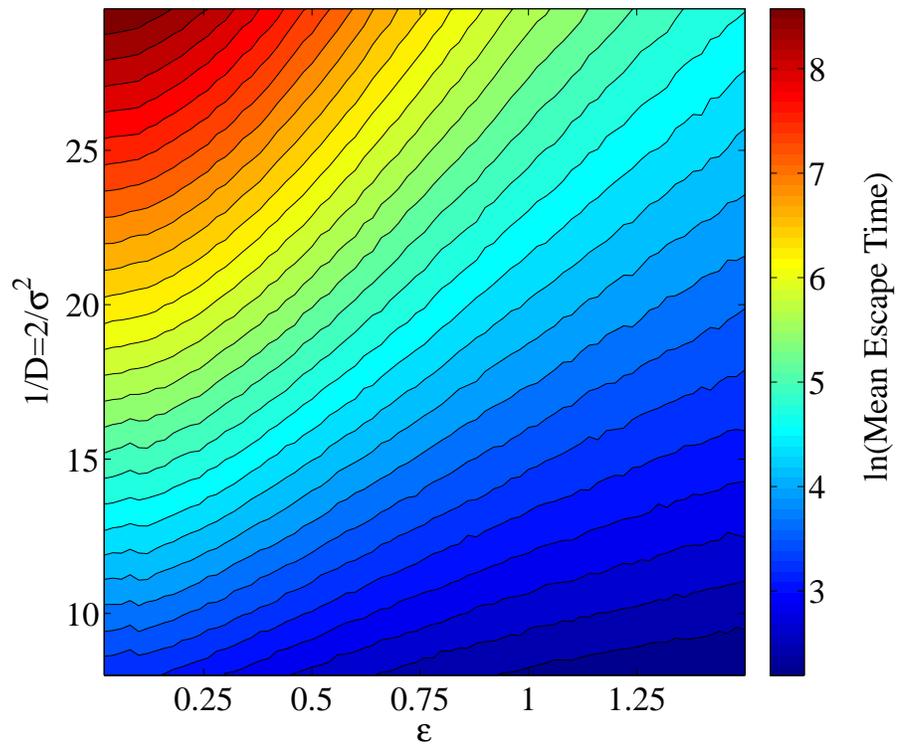}
\caption{\label{fig:MES_contour}Contours of numerically computed mean escape
  times (plotted as the natural log of the mean escape time) as a function of
  $\epsilon$ and $1/D$ for $10,000$ particles.}
\end{center}
\end{figure}
\clearpage
\begin{figure}[h!]
\begin{center}
\begin{minipage}{0.98\linewidth}
\includegraphics[width=12.5cm]{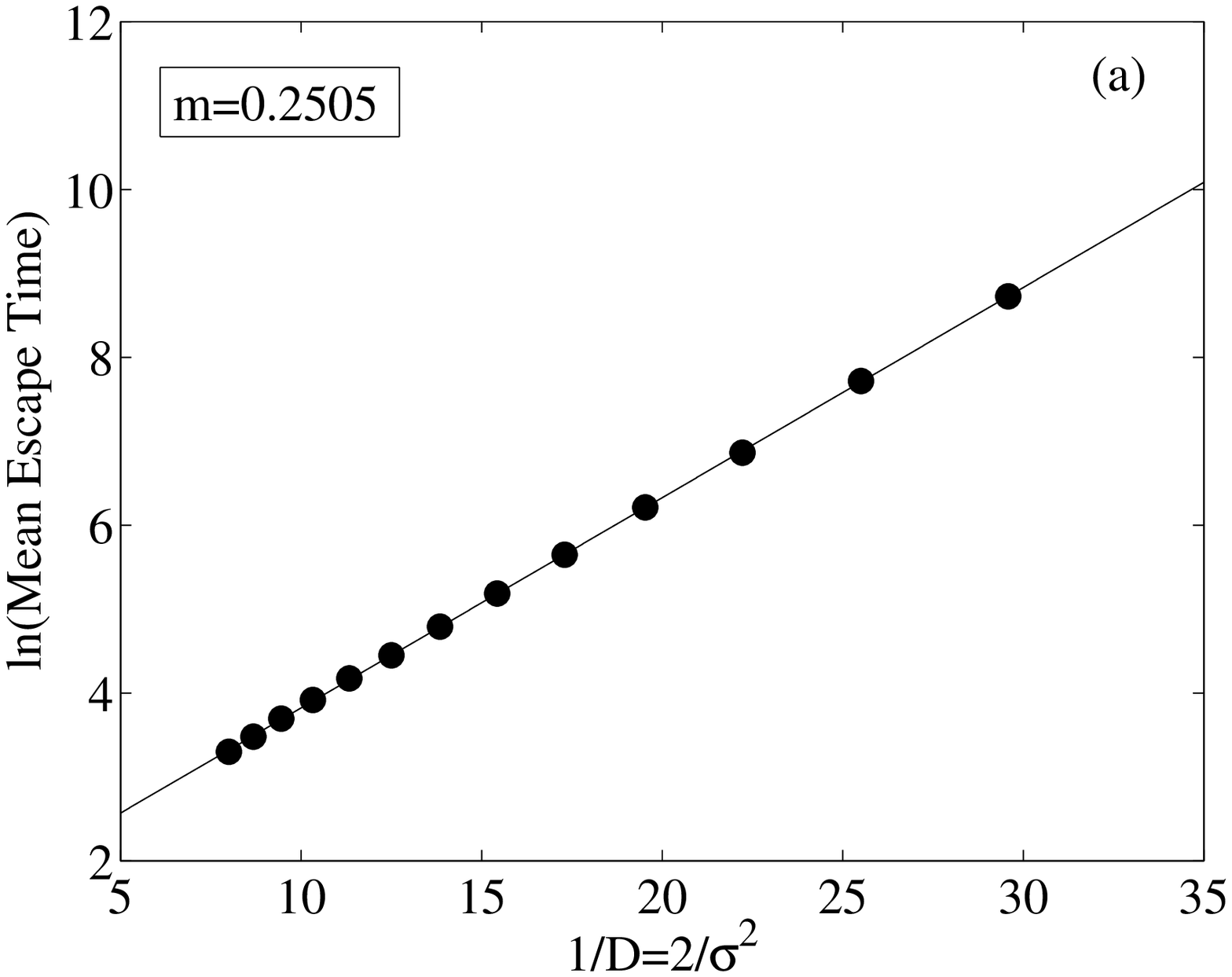}
\end{minipage}\\
\begin{minipage}{0.98\linewidth}
\includegraphics[width=12.5cm]{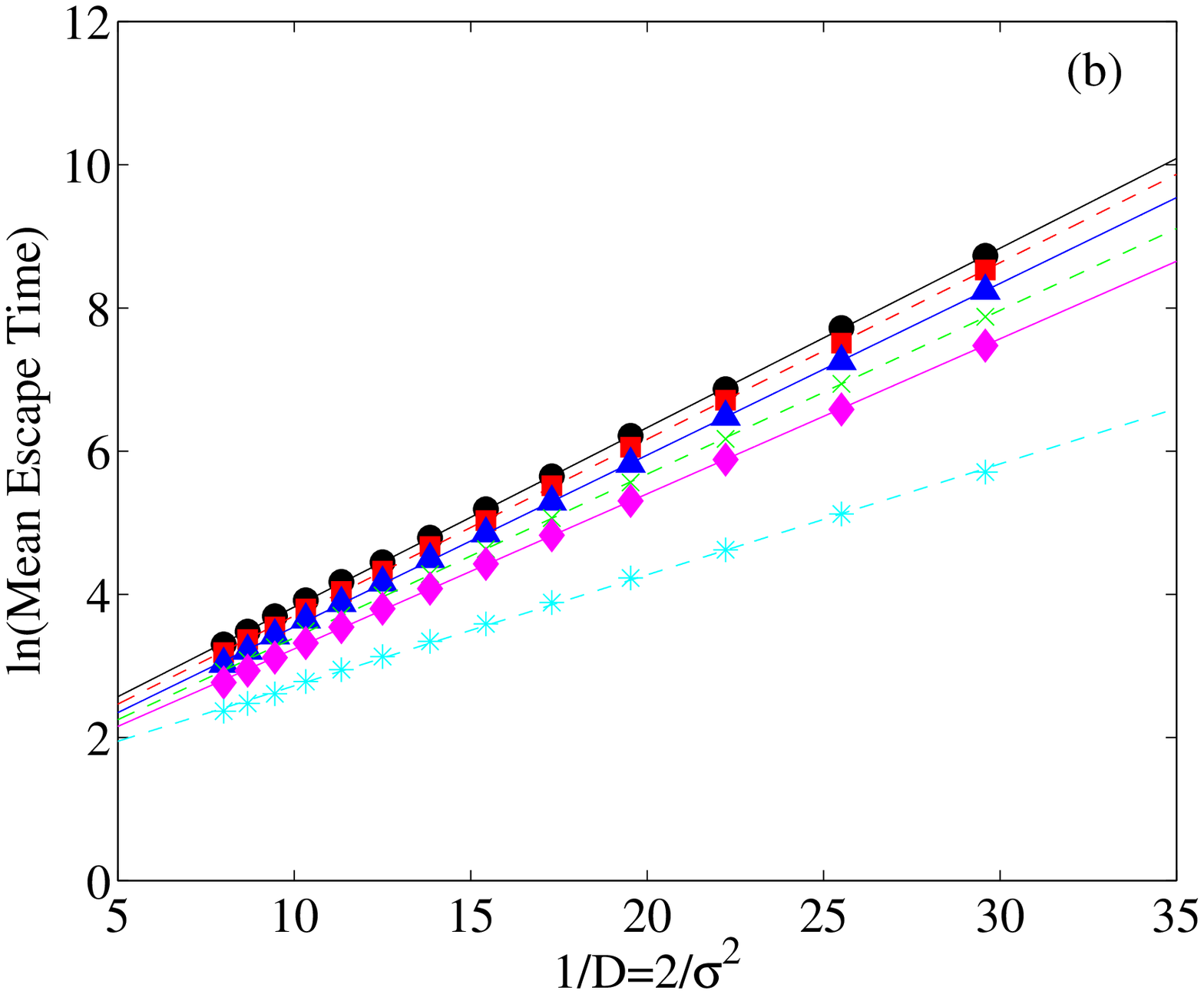}
\end{minipage}
\caption{\label{fig:MES_vs_noise}(a) Vertical slice of
Fig.~\ref{fig:MES_contour} taken at $\epsilon = 0.1$.  (b) Vertical slices of Fig.~\ref{fig:MES_contour} taken at $\epsilon = 0.1$
(black, ``circle'' markers), $\epsilon = 0.2$
(red, ``square'' markers), $\epsilon = 0.3$ (blue, ``triangle'' markers),
$\epsilon = 0.4$ (green, ``cross'' markers), $\epsilon = 0.5$
(magenta, ``diamond'' markers), and $\epsilon = 1.0$ (cyan, ``asterisk''
markers).  The data points in both (a) and (b) are overlaid by a line of best
fit.}
\end{center}
\end{figure}
\clearpage
\begin{figure}[h!]
\begin{center}
\begin{minipage}{0.49\linewidth}
\includegraphics[width=6.25cm]{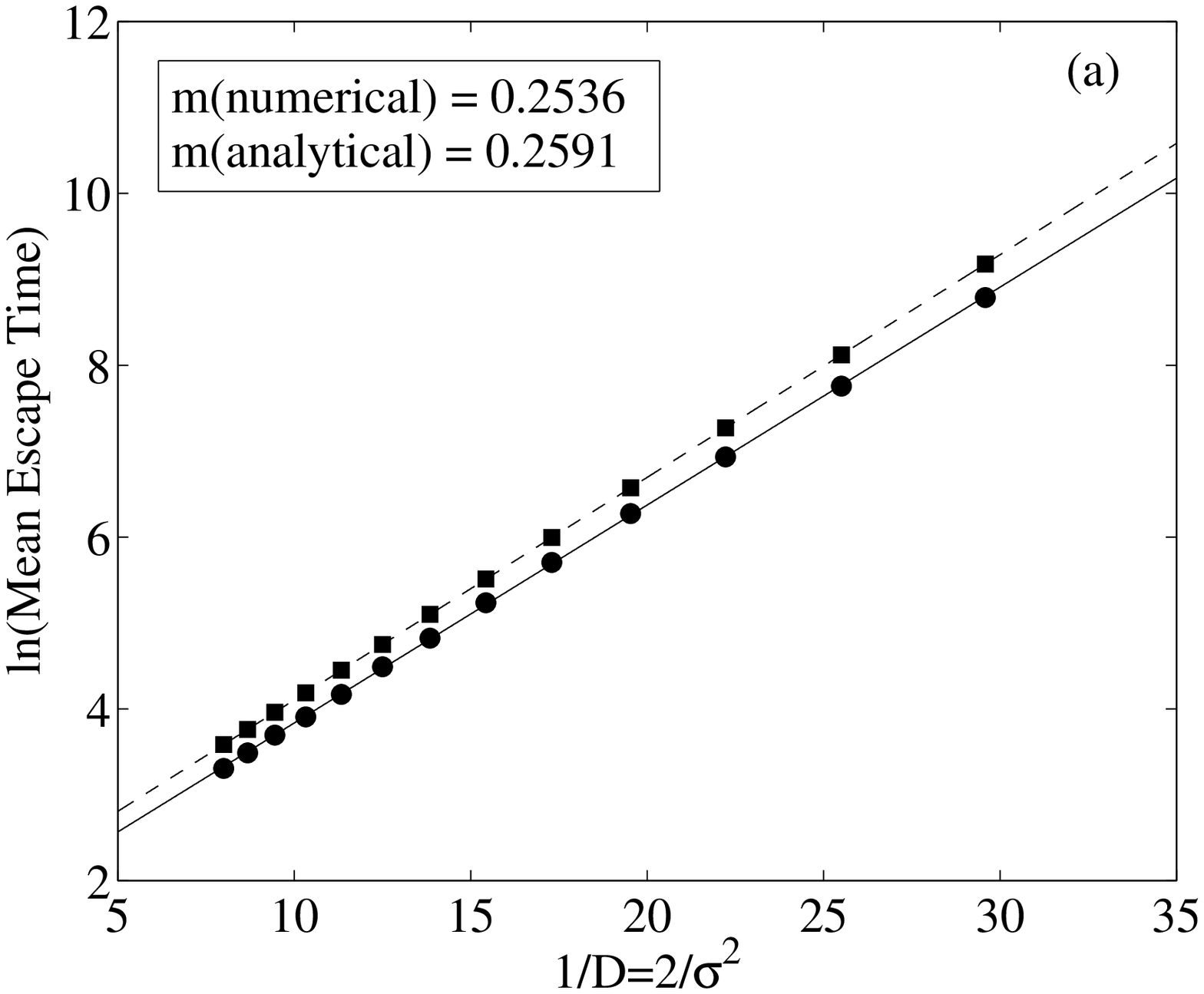}
\end{minipage}
\begin{minipage}{0.49\linewidth}
\includegraphics[width=6.25cm]{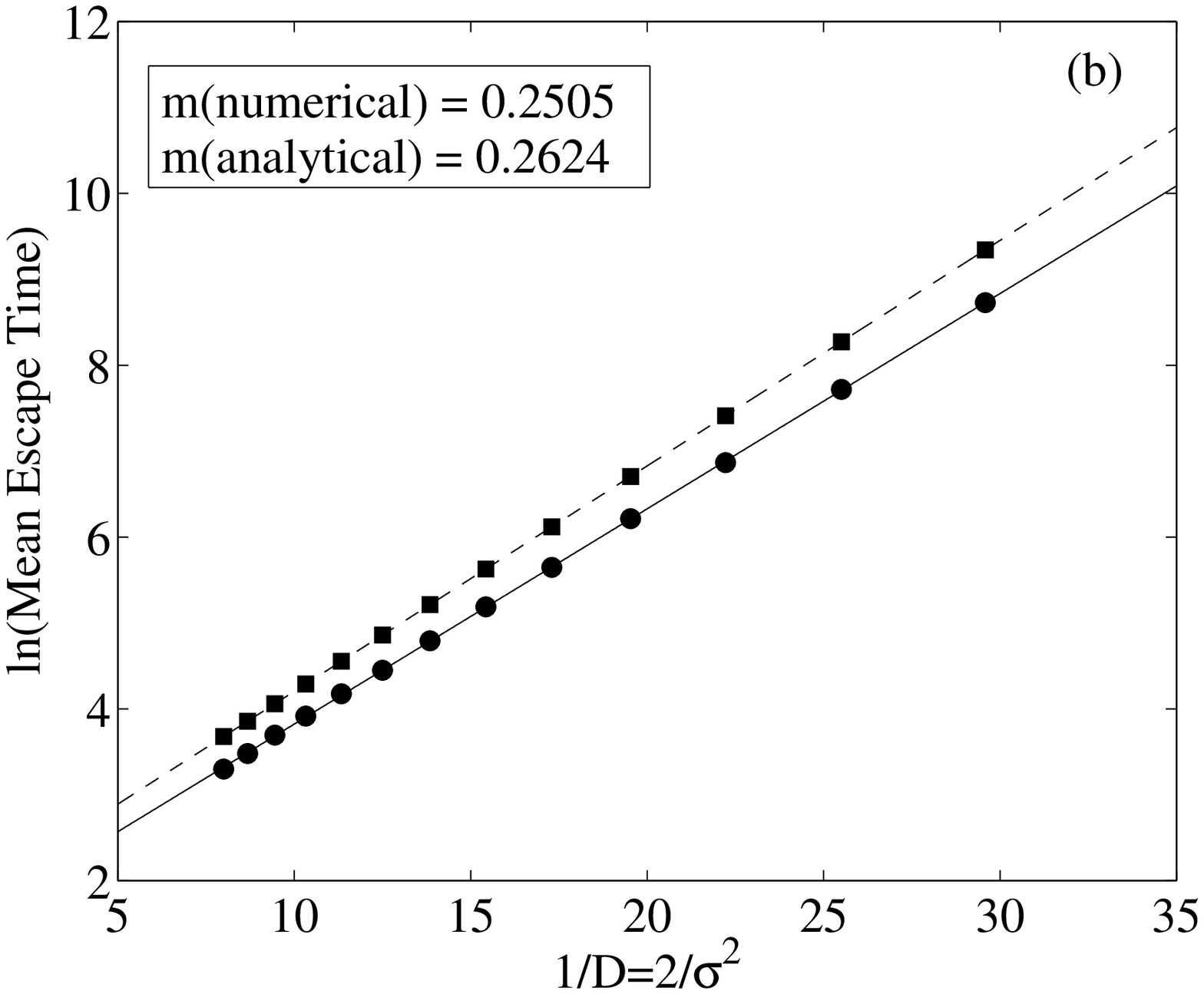}
\end{minipage}\\
\begin{minipage}{0.49\linewidth}
\includegraphics[width=6.25cm]{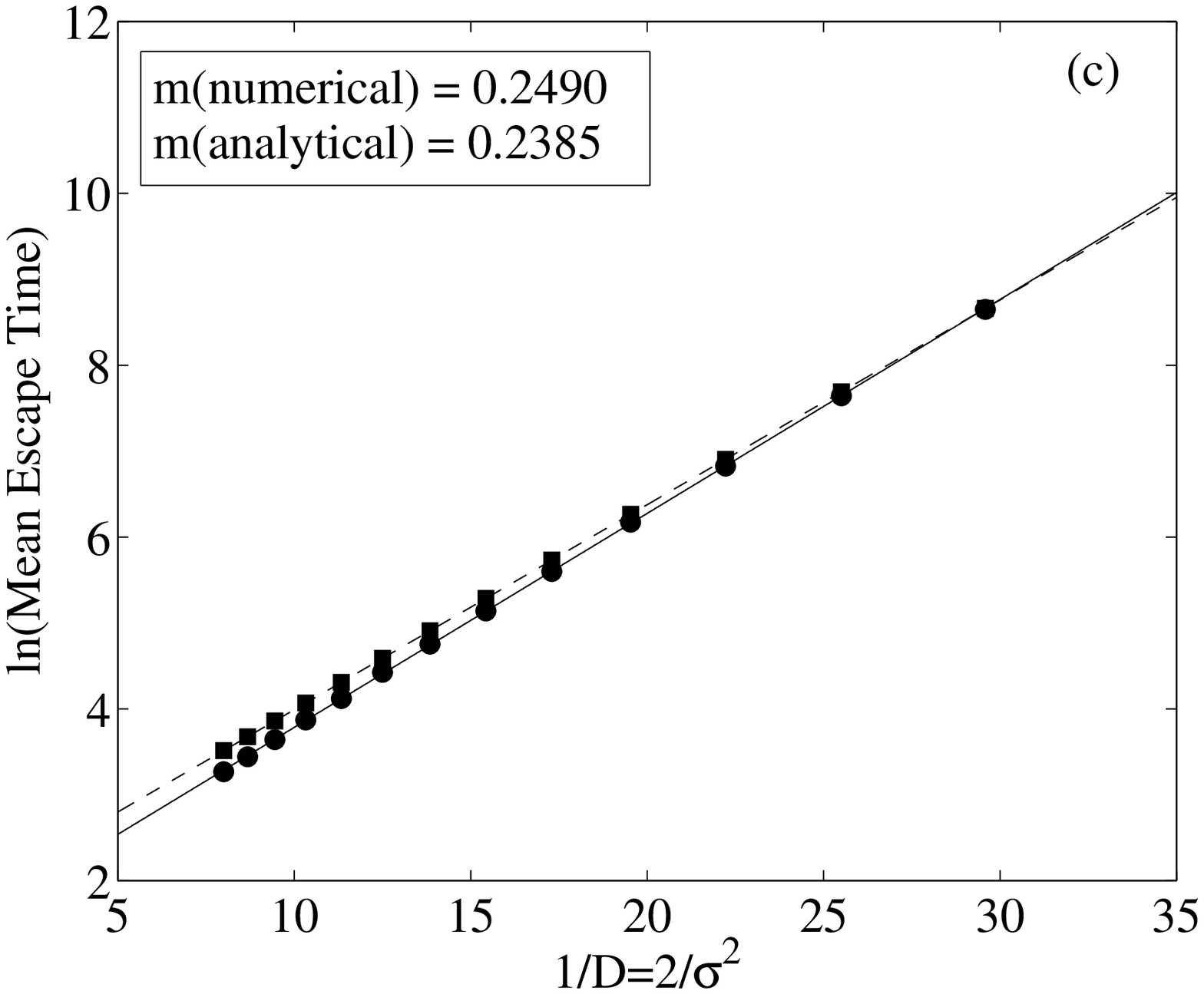}
\end{minipage}
\begin{minipage}{0.49\linewidth}
\includegraphics[width=6.25cm]{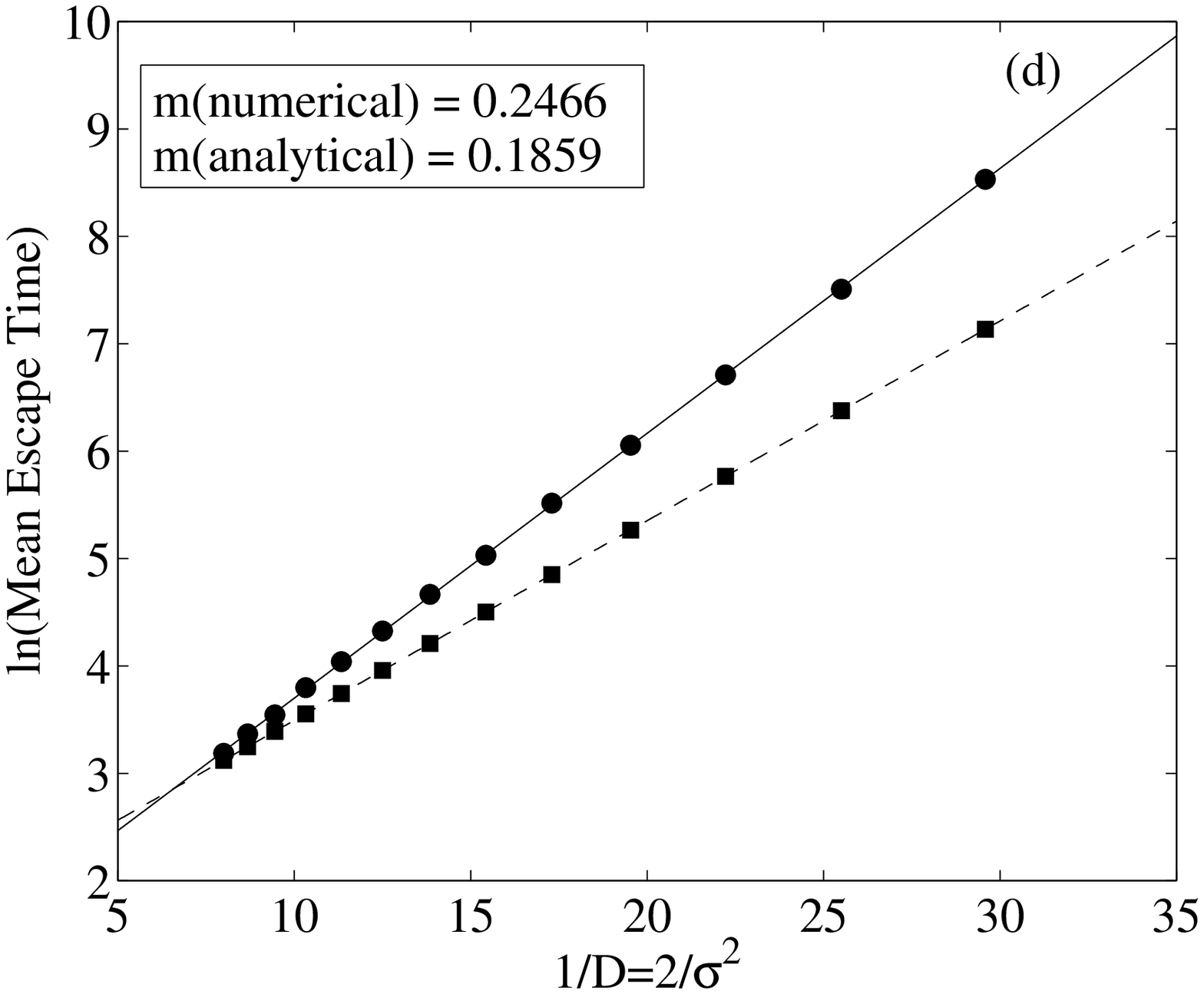}
\end{minipage}
\caption{\label{fig:MES_vs_noise_num_and_CM5}Comparison of numerical
data (``circle'' markers) and linear fit (solid line) with analytical data (``square'' markers) and
linear fit (dashed line) of the natural
log of the mean escape
time as a function of $1/D$ for (a) $\epsilon =0.02$, (b) $\epsilon =0.1$,
(c) $\epsilon =0.14$, and (d) $\epsilon =0.2$.}
\end{center}
\end{figure}
\clearpage
\begin{figure}[h!]
\begin{center}
\begin{minipage}{0.98\linewidth}
\includegraphics[width=12cm]{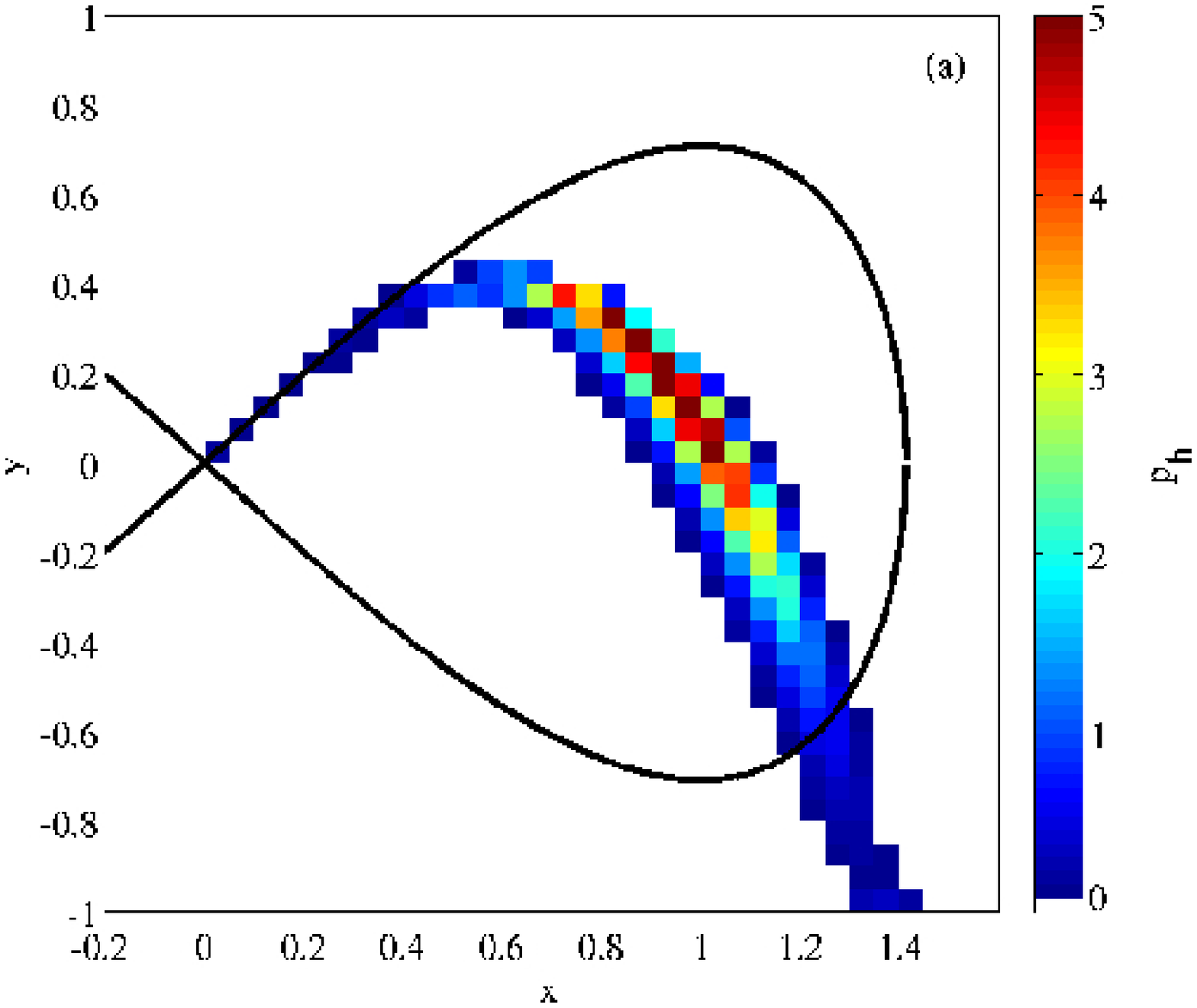}
\end{minipage}\\
\begin{minipage}{0.98\linewidth}
\includegraphics[width=12cm]{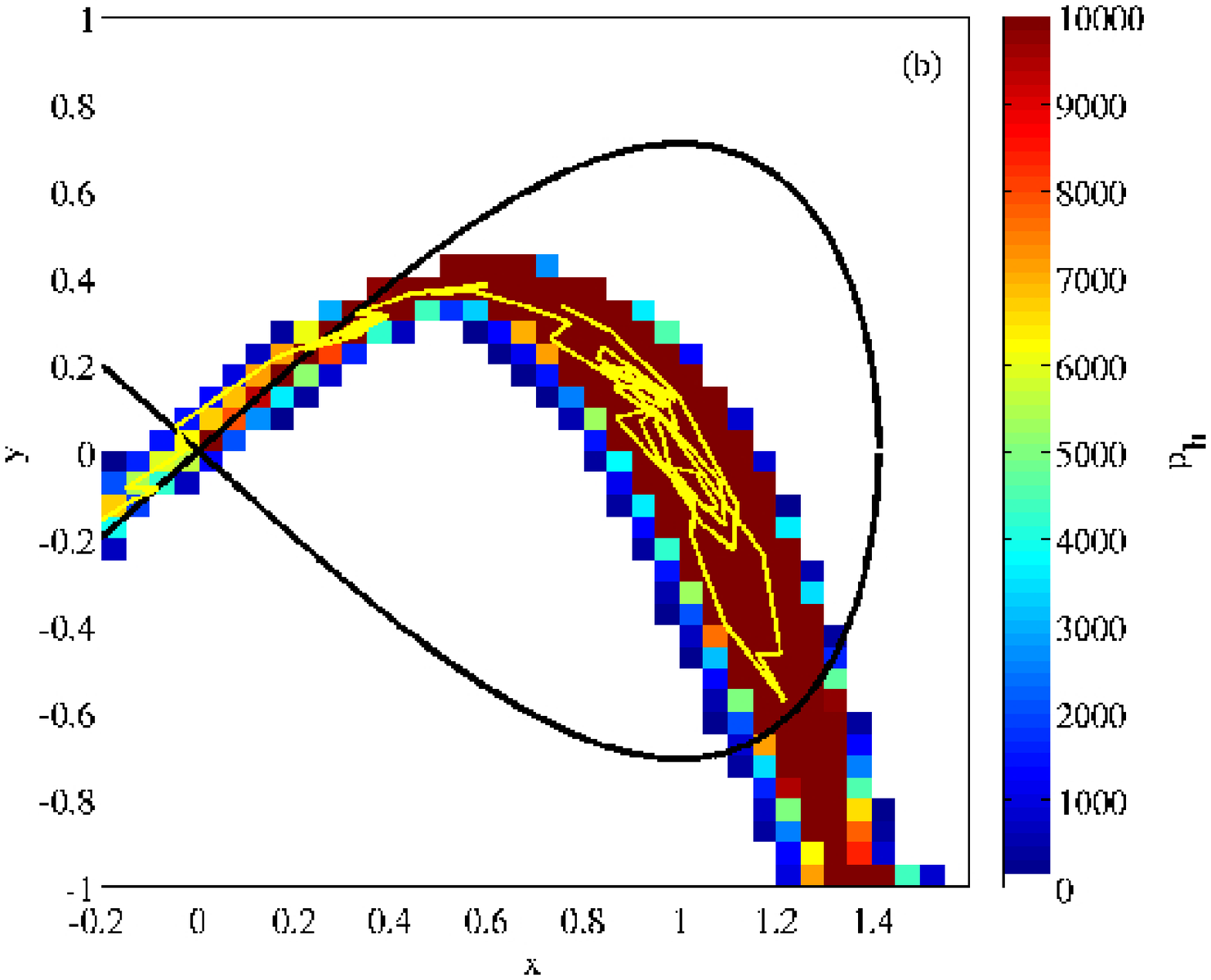}
\end{minipage}
\caption{\label{fig:hist_sig3_eps1_p10000}(a) Histogram of
  escape path prehistory (for $t=200$ of prehistory) of $10,000$ particles with $\epsilon=0.1$ and
  $\sigma = 0.3$.  The color-bar values have been normalized by $10^5$, and the
  threshold $0$ value is about $9000$.  (b) Same as Fig.~\ref{fig:hist_sig3_eps1_p10000}(a), but with adjusted
color-bar values, and including one particular particle's escape path prehistory (showing only
$t=50$ of prehistory).}
\end{center}
\end{figure}
\clearpage
\begin{figure}[h!]
\begin{center}
\begin{minipage}{0.98\linewidth}
\includegraphics[width=12cm]{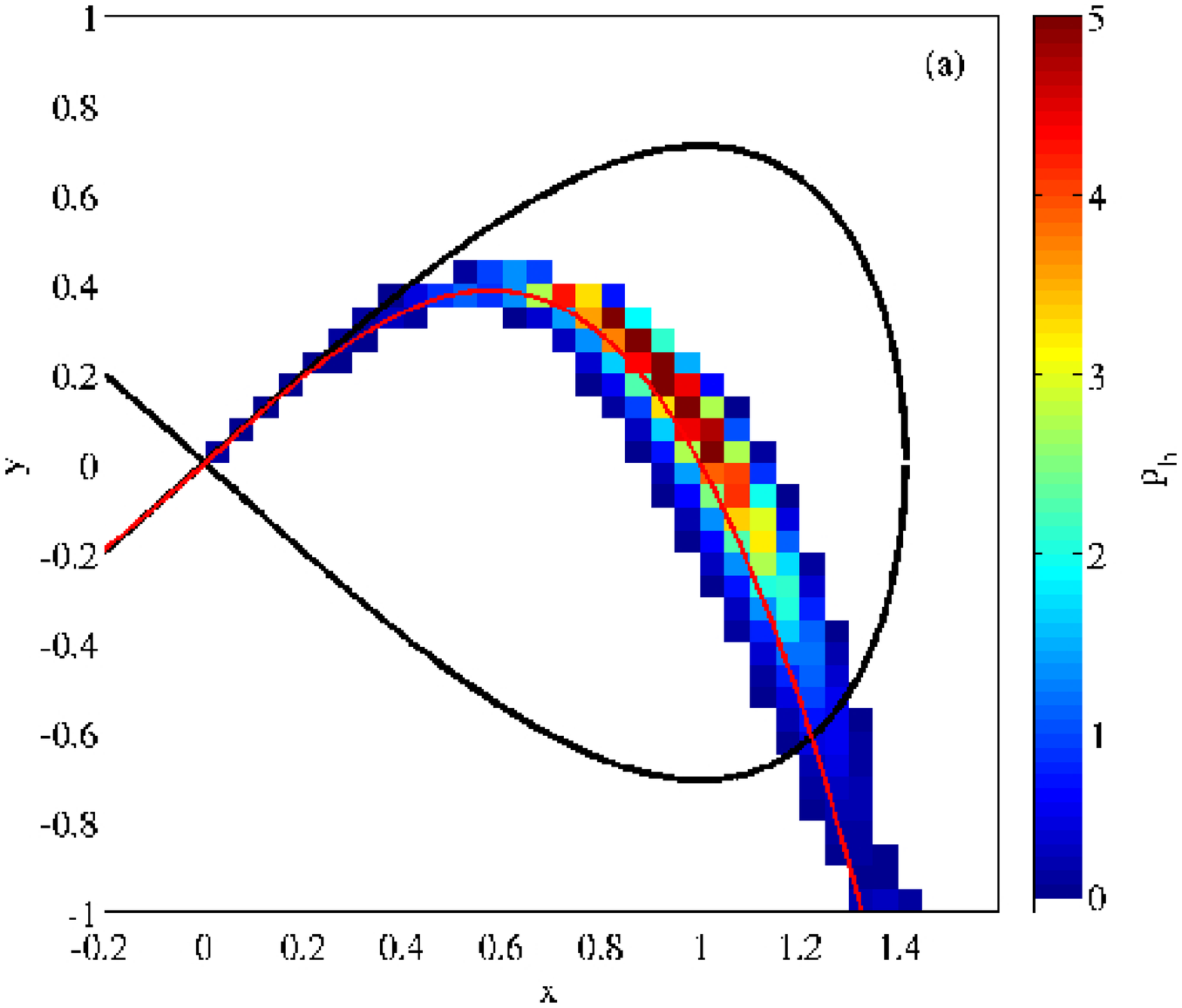}
\end{minipage}\\
\begin{minipage}{0.98\linewidth}
\includegraphics[width=12cm]{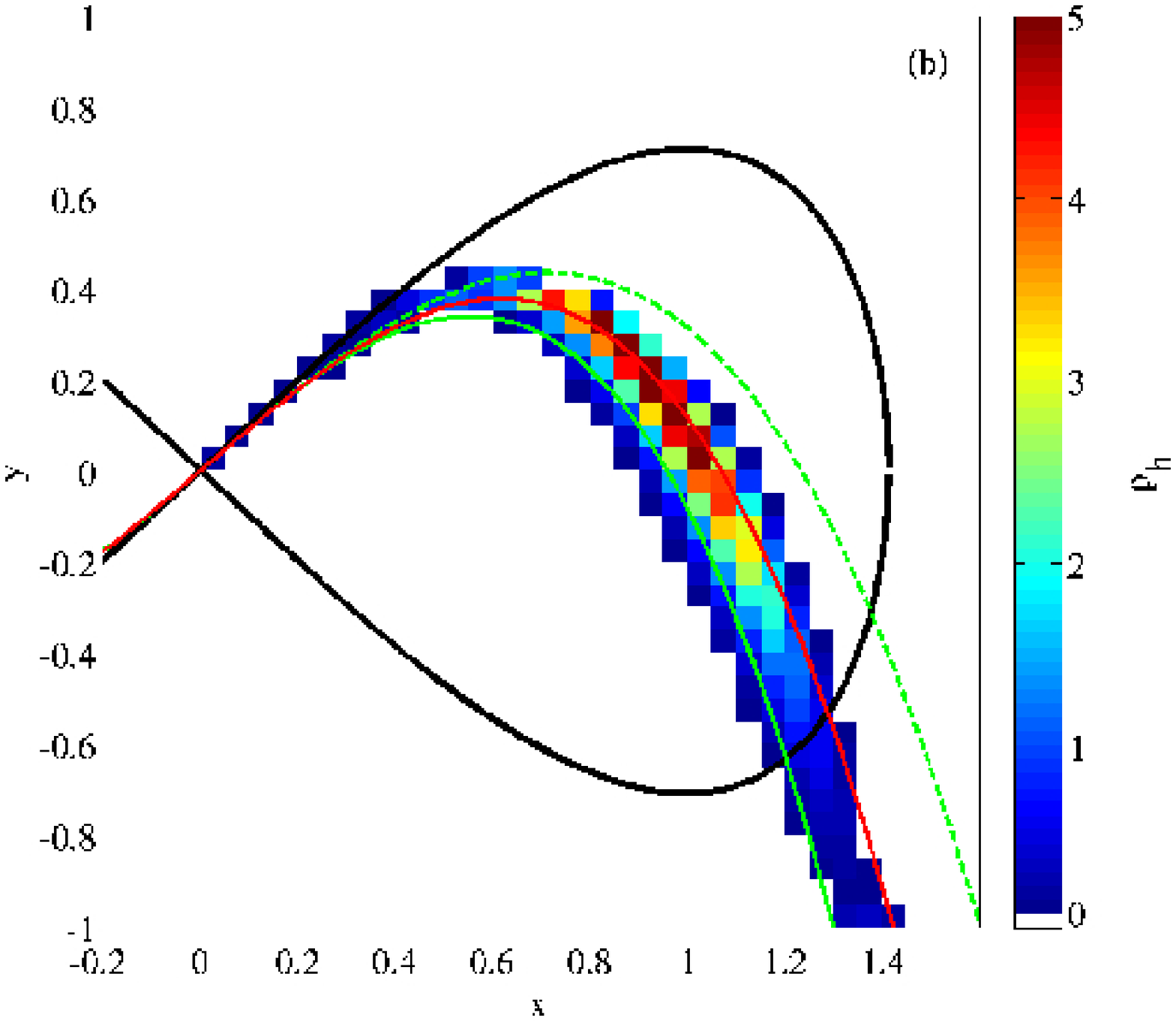}
\end{minipage}
\caption{\label{fig:hist_sig3_eps1_p10000_man}Escape path prehistory histogram of
Fig.~\ref{fig:hist_sig3_eps1_p10000}(a) overlaid with (a) the graph of the
slow manifold equation, and
(b) the graphs of the third-order (solid, green line), fourth-order (dashed, green
line), and fifth-order (solid, red line) center
manifold equations given by Eq.~(\ref{e:h3}).  For both Figs.~\ref{fig:hist_sig3_eps1_p10000_man}(a)
and~\ref{fig:hist_sig3_eps1_p10000_man}(b), the color-bar values have been normalized by $10^5$, and the threshold $0$ value is about $9000$.}
\end{center}
\end{figure}
\clearpage
\begin{figure}[h!]
\begin{center}
\begin{minipage}{0.98\linewidth}
\includegraphics[width=12.5cm]{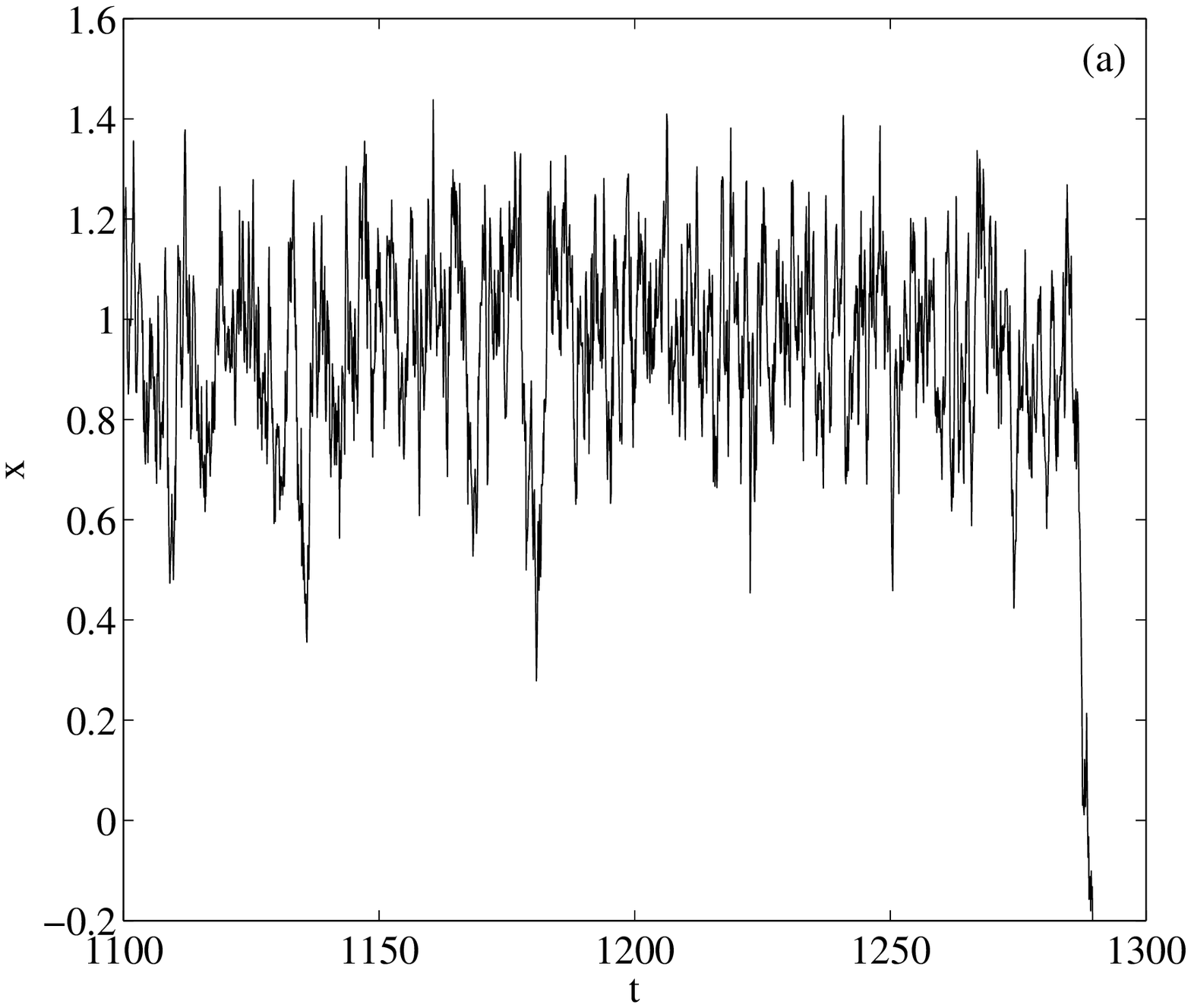}
\end{minipage}\\
\begin{minipage}{0.98\linewidth}
\includegraphics[width=12.5cm]{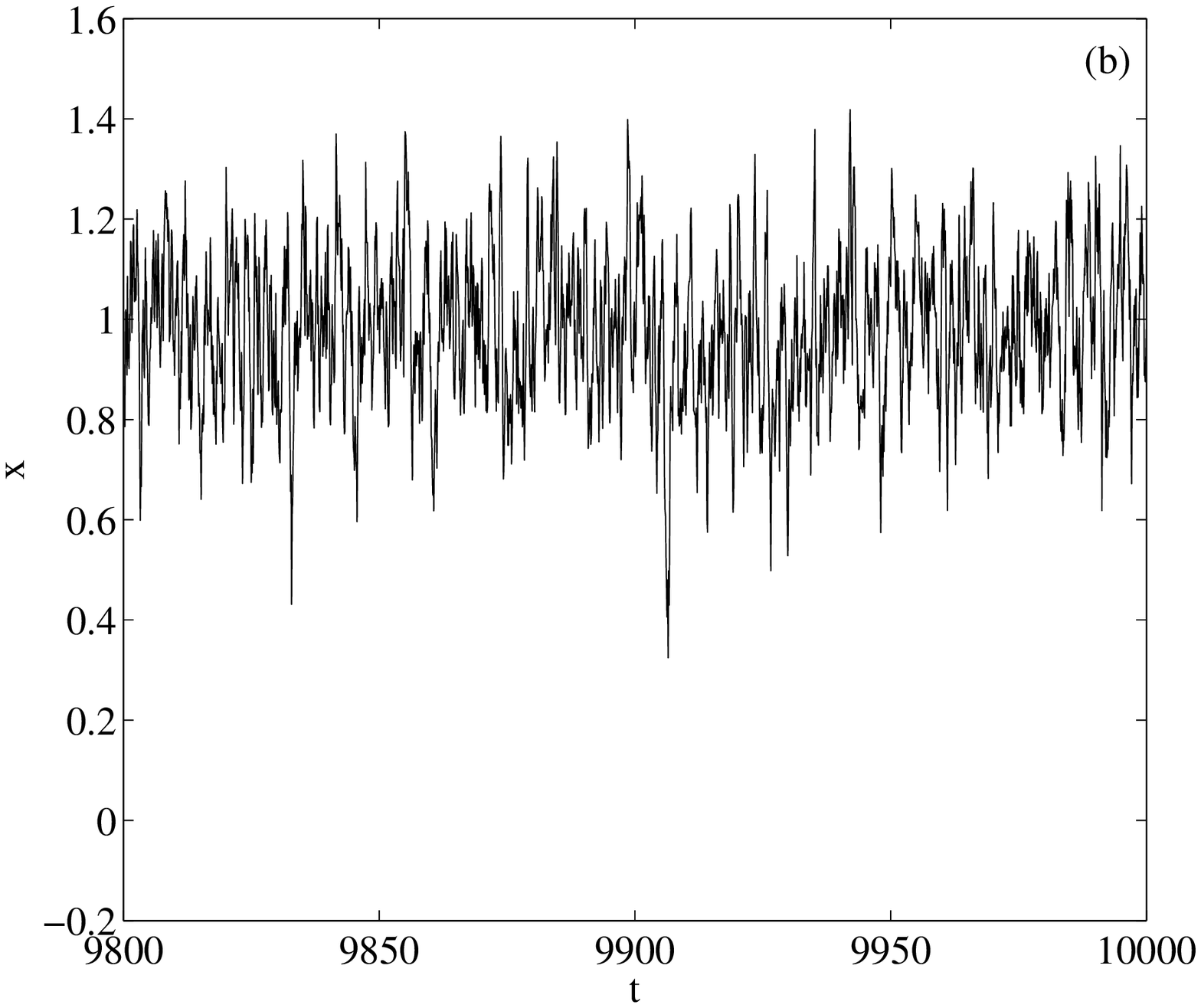}
\end{minipage}
\caption{\label{fig:x_vs_t}Location of a particle ($x$ coordinate) as a
  function of time $t$ (a) without control, and (b)
with control.}
\end{center}
\end{figure}
\clearpage
\begin{figure}[h!]
\begin{center}
\begin{minipage}{0.98\linewidth}
\begin{center}
\includegraphics[width=12.5cm]{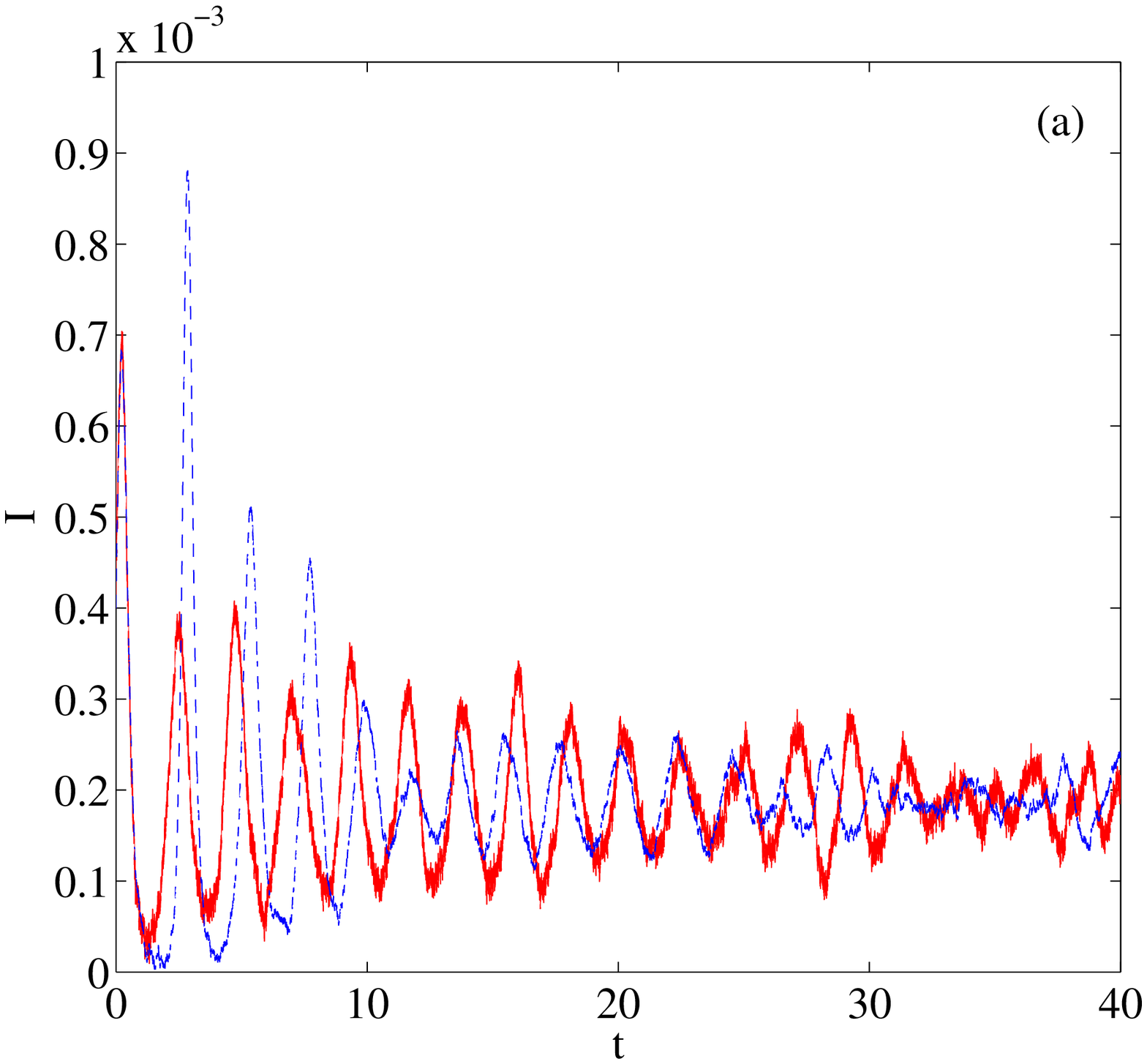}
\end{center}
\end{minipage}\\
\begin{minipage}{0.98\linewidth}
\begin{center}
\includegraphics[width=12.5cm]{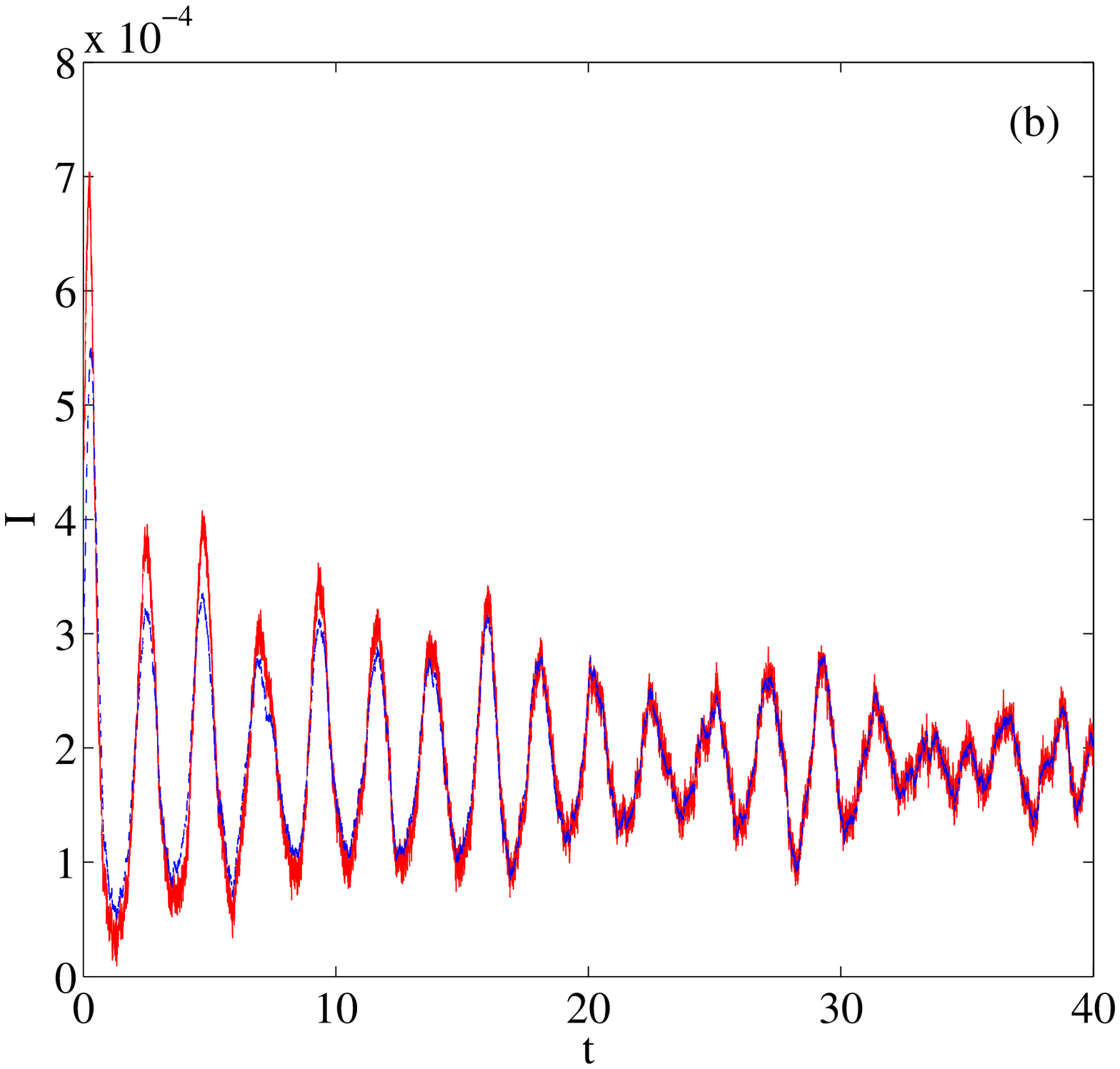}
\end{center}
\end{minipage}
\caption{\label{fig:SEIR}Time series of the fraction of the
  population that is infected with a disease, $I$, computed using the
  complete, stochastic system of equations of the SEIR model (red, solid
  line), and (a) computed using the reduced system of equations of the SEIR
  model that is based on the deterministic center manifold with a
  ``na{\"i}ve'' replacement of the noise terms (blue, dashed line), and (b) computed using the reduced system of equations of the SEIR
  model that is found using the stochastic normal form coordinate transform
  (blue, dashed line). }
\end{center}
\end{figure}
\end{document}